\documentclass[
reprint,
 amsmath,amssymb,
 aps,
superscriptaddress]{revtex4-1}

\usepackage{graphicx}% Include figure files
\usepackage{dcolumn}% Align table columns on decimal point
\usepackage{bm}% bold math
\usepackage{hyperref}% add hypertext capabilities
\usepackage[mathlines]{lineno}% Enable numbering of text and display math
\usepackage[table,xcdraw]{xcolor}

\newcommand*\rot{\rotatebox{90}}
\newcommand{\ket}[1]{|#1\rangle}

\newcommand{\sech}[1]{\textrm{sech}\left(#1\right)}

\newcommand{\kk}{\mathbf{k}}
\newcommand{\pp}{\mathbf{p}}

\newcommand{\JJ}{\mathbf{J}}

\newcommand{\AAA}{\mathbf{A}}

\newcommand{\nin}{\noindent}

\newcommand{\ppi}{\bm{\pi}_{\kk}}

\newcommand{\ssigma}{\bm{\sigma}}

\newcommand{\eb}{\epsilon^{b}_{\kk}}

\usepackage{float}
\usepackage{blindtext}
\usepackage{rotating}
\usepackage{hhline}
\usepackage{multirow}
\usepackage{comment}
\usepackage{amsmath}
\usepackage{tabularx}
\usepackage{braket}
\usepackage{epstopdf}

\begin{document}

\preprint{ArXiV}

\title{Role of Anisotropy in Nonlinear Harmonic Generation Across TMD Monolayers}

\author{David N. Carvalho}
\email{David.Carvalho@su.se. The work done by the author has been performed exclusively while affiliated in $^{2}$. }
\affiliation{Nordita, KTH Royal Institute of Technology and Stockholm University, Roslagstullsbacken 23, SE-106 91 Stockholm, Sweden}
\affiliation{School of Engineering and Physical Sciences, Heriot-Watt University, EH14 4AS Edinburgh, UK \\}
%\altaffiliation[Also at]{email: dc8@hw.ac.uk}
\author{Fabio Biancalana}
\email{F.Biancalana@hw.ac.uk}
\affiliation{School of Engineering and Physical Sciences, Heriot-Watt University, EH14 4AS Edinburgh, UK \\}

\date{\today} 

\begin{abstract}
Recent techniques have allowed transition metal dichalcogenides (TMD) monolayers to be grown and adequately characterised. Of particular interest, their nonlinear optical response presents many promising opportunities for future nanophotonic devices and technology. The dispersion of the carriers is trigonally-warped, leading to an anisotropic Fermi surface for low-lying states.
In this paper, the effects of such a deformation on the nonlinear harmonic generation are studied by considering a tight-binding model expanded up to third order in $\kk \cdot \pp$. By solving exactly the free-carrier dynamics of the carriers when interacting with intense and ultrashort pulses of light, we predict the photo-generated current in a nonperturbative way and study its harmonic composition. We find frequency and amplitude modulation of the nonlinear current in quadratic and cubic models. Furthermore, we demonstrate anisotropy-induced modulation of the intensity of higher-order harmonics and the existence of harmonic crossovers, depending on the incident light polarisation. The methodology presented in this paper may be applied to any general effective two-band model and offer a pathway to identify signatures of electronic features in optical output. 

%\begin{description}
%\item[PACS numbers]
%May be entered using the \verb+\pacs{#1}+ command.
%\end{description}

\end{abstract}

\pacs{Valid PACS appear here}% PACS, the Physics and Astronomy Classification Scheme.


\maketitle

%\tableofcontents

\section{Introduction}
\label{sec:intro}

\nin An ample range of semiconductor crystals -- \emph{transition metal dichalcogenides} (TMD) -- has recently sparked the attention of the research community for reliably exhibiting robust optical and electronic properties when grown in the form of ultrathin, two-dimensional monolayers. Presently, most prominent and widely researched TMDs include disulphides (MoS$_2$), diselenides (WSe$_2$,  MoSe$_2$, WSe$_2$) as well as ditellurides (WTe$_2$, MoTe$_2$). 

\nin These structures admit direct bandgaps around the near-infrared/visible range when grown in monolayers \cite{Kolobov2016, Singh2018}. Various ab-initio calculations support the notion that electron-hole symmetry is generally broken \cite{Kormanyos2013} and trigonal warping effects, responsible for distortions in the otherwise isotropic energy dispersion, have also been observed and studied \cite{Saynatjoki2017}. 

\nin These systems provide an interesting platform to study new physics. Due to their orbital motion, electron states split according to their spin through spin-orbit coupling (SOC), giving rise to important sub-band energy gaps. Giant spin splittings, in the order tenths of eV, have been reported \cite{Zhu2011}, making them suitable candidates to probe spintronic effects. Valleytronics, in analogy to spintronics, has also been gaining ground as to provide a platform to study many novel phenomena associated to the valley degree of freedom. It has been suggested that, within a low-momentum approximation, spin-valley-locked dynamics and valley-dependent electromagnetic response \cite{valleytronicsin2d2016, Yao2014} through optical selection rules for specific light field polarisations may be achieved due to strong SOC \cite{Xiao2012} in these crystals. \\
\nin Excitonic effects, manifested through the existence of both dark and bright solitons, biexcitons and trions \cite{Zheng2018}, lead, generally speaking, to enhancement of many optical properties \cite{Liu2016, Yu2014}. Their exact characterisation in TMDs is challenging to quantify, since they depend strongly on the two-dimensional dielectric environment. \\
\nin Strong optical nonlinearities have been established through a myriad of methods. Second and third-order susceptibility have been both theoretically and experimentally studied in a wide range of TMD flakes \cite{Marini2018, Khorasani2018}, showing different responses. Z scans and pump-probe experiments have established a remarkably strong layer-dependent nonlinear response in MoS$_2$ samples and also estimated relaxation rates \cite{Wang2018}. Nanophotonic devices and applications to enhance light-matter interactions in these materials have also been proposed \cite{Xia2014, Singh2018}, as well as heterostructures e.g. by depositing graphene on TMD monolayers \cite{Gmitra2016}. \\
\nin Of relevance to harmonic generation, these crystals lack a centre of inversion. Second-order optical nonlinearities are thus expected. Rather strong signatures of this have already been observed in MoS$_2$ monolayers \cite{Saynatjoki2017}.  \\
\nin Models containing as many as eleven bands have been proposed to model the dispersion across the entire Brillouin zone (BZ). However, quasi-degenerate perturbation theory (also known as L\"owdin partitioning) allows reduced models taking into account two effective bands to be obtained in satisfactory momentum ranges, providing more tractable theoretical machinery \cite{Liu2013, Rostami2013}. With the aid of phenomenological energy parameters, the model given in \cite{Liu2013} reproduces experimental data of various TMDs very well and offers a very adequate Hamiltonian to understand the role of trigonal warping, electron-hole asymmetry, SOC coupling and gap in the generation of current in the sample. \\
\nin In this paper, we focus on nonlinear signatures in the light-matter interactions in these materials, in the hope of advancing the understanding of how these features, captured in a suitable $\kk \cdot \pp$ expansion, influence the optical response. The theoretical machinery introduced may be used to compute the output current in a non-nonperturbative fashion i.e. without expanding in the electromagnetic field, for a general system which can be suitably described by an effective two-band model. \\

\nin This paper is organised as follows: in Sec. \ref{sec:quasiparticle_characterisation}, the quasiparticle excitations in TMD crystals are briefly outlined. In Sec. \ref{sec:lattice}, their lattice arrangement is explained, focusing on its symmetries and lack of centrosymmetry when grown in monolayers. In Sec. \ref{sec:quasiparticle_dynamics}, we outline how the coupling of the electromagnetic field is performed and derive the instantaneous eigenstates, respective energy dispersion and electric dipole moment of an effective two-band model. In Sec. \ref{sec:generalised_DBEs}, we provide a set of dynamical equations -- the \emph{Dirac-Bloch Equations} -- which describes the evolution of the populations and coherences of an effective two-level system. With these quantities, the output current and its respective spectrum may be obtained nonperturbatively. This procedure is outlined in Sec. \ref{sec:current}. With the machinery laid out, we apply the model to a Mo$S_2$ monolayer and show the output current and spectra in Sec. \ref{sec:simulations}. In particular, we explicitly show the role of inversion-breaking terms in the generation of even harmonics in Sec. \ref{sec:cumsim} and how the light polarisation angle modulates and lead to crossover between harmonic order intensities in Sec. \ref{sec:directionalsim}. In Sec. \ref{sec:conclusions}, we summarise our findings, comment on the validity of the model and offer prospective insights into their applicability.

\section{Quasiparticle characterisation}
\label{sec:quasiparticle_characterisation}

\subsection{Lattice and underlying symmetries}
\label{sec:lattice}

\begin{figure}[ht]
\centering
3\includegraphics[width=8.45cm]{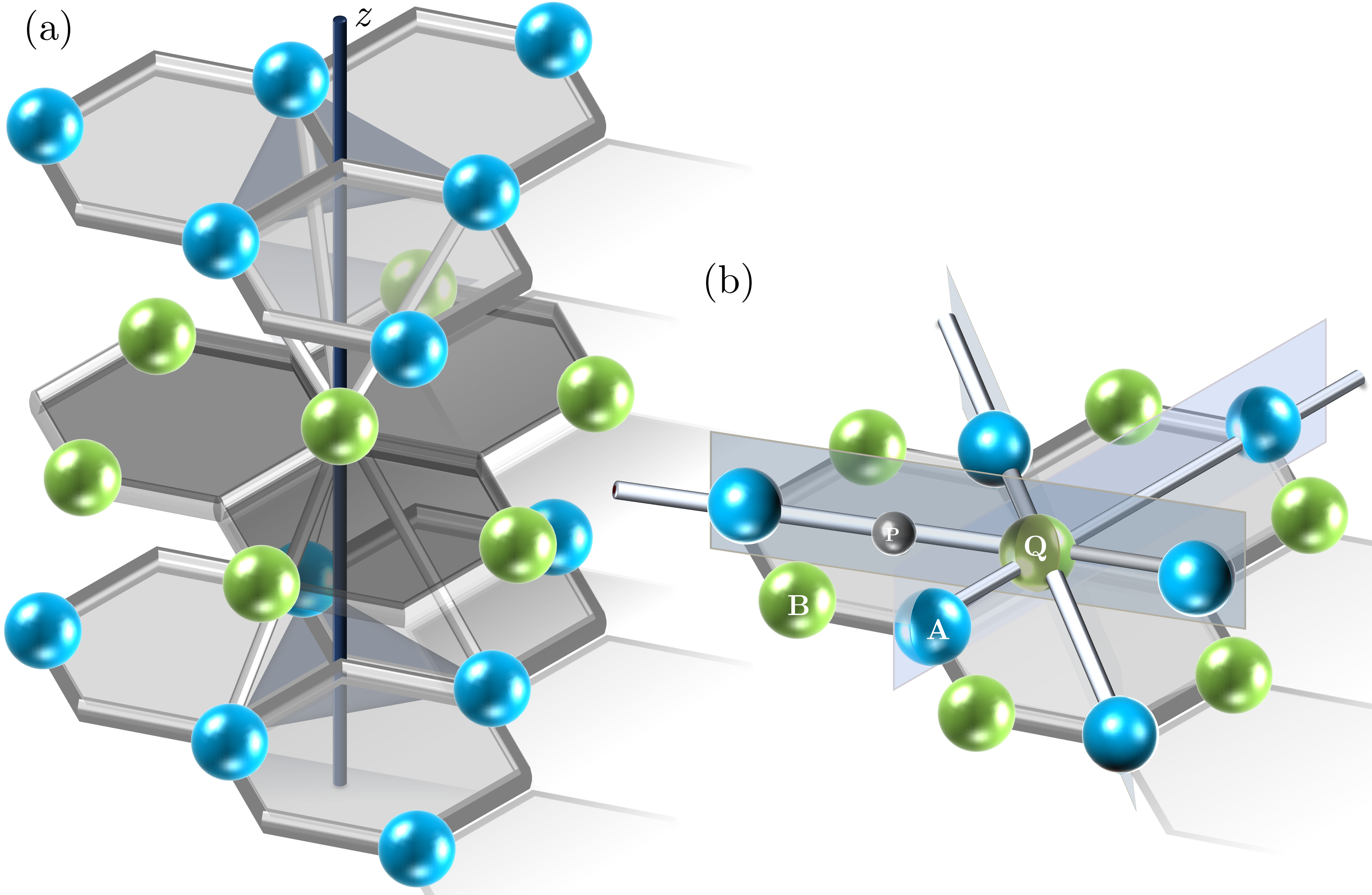}
\caption{Lattice structure of a TMD monolayer in the trigonal prismatic arrangement. (a) The transition metal layer (green atoms on dark gray plane) is interposed between two identical layers of a chalcogen (blue atoms on light gray planes) and spatially stacked so that the metal atoms are coordinated with the three nearest chalcogen atoms on either respective layer, represented by the white bonds. The vertical axis (black) and the horizontal metal-occupied plane generate three 3-fold rotations, one reflection and two improper rotations as symmetry operations. (b) Top-view of the monolayer, with both chalcogen-occupied layers superimposed. Both species occupy alternate triangular sublattices $\mathbf{A}$ and $\mathbf{B}$. The point group symmetry of the monolayer is completely  determined by including three further reflections about the three gray axes and three reflections about the vertical  planes. The only hypothetical monolayer centres of inversion, arbitrary taken at the points $\mathbf{P}$ and $\mathbf{Q}$, are shown. As explained in the main text, inversion through these points can never yield a symmetry.} 
\label{fig1}
\end{figure}

\nin TMDs are structures of the form MX$_{2}$, composed of a layer of a transition metal M interposed between two layers of a chalcogen X and with a typical thickness of around $6-7$\AA. Depending on the metal's group, their stacking arrangement can either be trigonal prismatic or octahedral. For group-VI metals, such as Mo and W, these three layers tend to stack in the former arrangement. This can be seen in Fig.~\ref{fig1}(a), where the two species, which bond covalently, occupy either one of the two possible triangular sublattices, labelled $\mathbf{A}$ and  $\mathbf{B}$, in a honeycomb lattice, effectively forming a monolayer if viewed from the top [Fig.~\ref{fig1}(b)]. Due to weak van der Waals interlayer interactions, multilayer or bulk structures can be engineered. Extensive and detailed analysis of the chemical bonding mechanisms behind TMDs may be found in \cite{Kolobov2016}.

\nin A TMD monolayer admits twelve symmetry operations, which are illustrated in Fig.~\ref{fig1}. These consist of three 3-fold rotations around the black axis in Fig.~\ref{fig1}(a), three 2-fold rotations around the axes laying on the horizontal plane, as shown in Fig.~\ref{fig1}(b), two 3-fold improper rotations (composed of a 3-fold rotation followed by a reflection on the metal-occupied plane) and three reflections on their respective vertical planes. However, this structure lacks an inversion centre. To see why, the only hypothetical inversion centres, the centre of an arbitrary hexagon (labelled $\mathbf{P}$) and a lattice point $\mathbf{Q}$ are shown in Fig.~\ref{fig1}(b). Inversion through $\mathbf{P}$ is not possible due to the alternate nature of each sublattice; as for $\mathbf{Q}$, only the lattice points of the same species are invertible (in this instance the metal species). This arrangement cannot therefore preserve inversion symmetry. The precise configuration of the TMD structure plays a role in the observation of this symmetry. Unlike monolayers, TMDs grown in bulk or multilayers of an even number of layers admit an inversion centre \cite{Kolobov2016}. \\

\subsection{Quasiparticle dynamics}
\label{sec:quasiparticle_dynamics}

\begin{figure}[ht]
\centering
\includegraphics[scale=0.2]{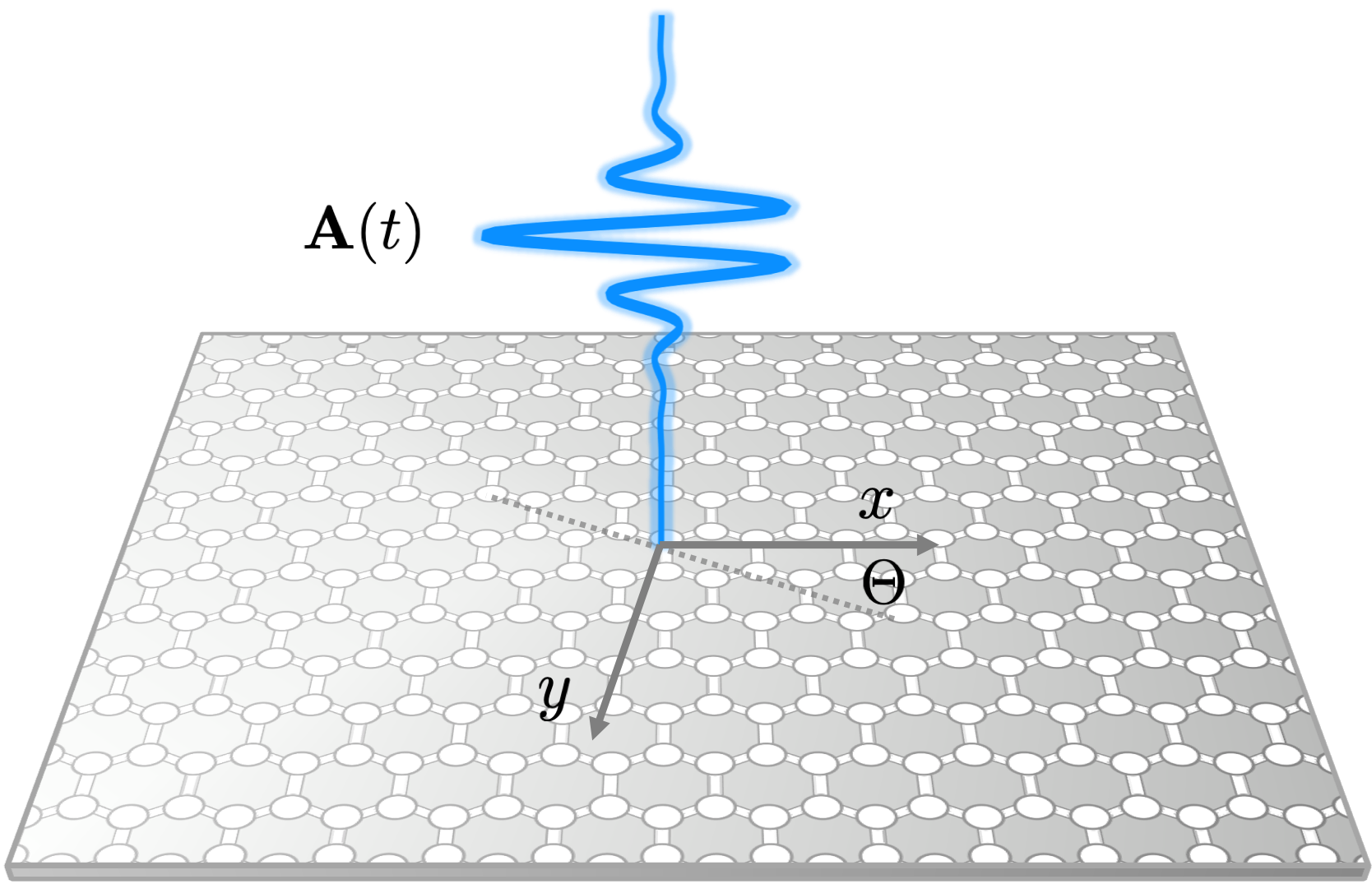}
\caption{Schematics of the geometry considered. The TMD crystal lies on the $x-y$ plane and is illuminated by a linearly-polarised pulse travelling perpendicularly to the monolayer. The respective vector potential oscillates on the plane at an angle $\Theta$ with respect to the lattice.}
\label{fig:light}
\end{figure}

\nin We assume normal incidence conditions, whereby light is assumed to move perpendicularly with respect to the monolayer. In order to study the effects of nonlinear ultrashort light-matter interactions, we choose the velocity gauge, wherein the electromagnetic field couples to the carriers via the minimal substitution: the in-plane carrier momentum $\kk = (k_{x}, k_{y}) \equiv |\kk|(\cos \phi_{\kk},\sin \phi_{\kk})$ is mapped to a time-dependent wavevector $\ppi(t)$, through the assignment $\kk \mapsto \ppi(t) \equiv \kk + e/(\hbar c) \AAA(t)$, where $\AAA(t)$ is the electromagnetic vector potential. We consider linearly polarised light and thus introduce a \emph{directional angle} $\Theta$, which sets the direction of oscillation of the electric field relative to the monolayer plane. A schematic depiction of the geometry considered may be found in Fig.~\ref{fig:light}. \\
\nin Consequently, one has $\textbf{A}(t) = (A_{x}(t), A_{y}(t)) \equiv  A(t) (\cos \Theta, \sin \Theta)$. The canonical momentum is now split in polars through $\ppi(t) = (\pi_{x}(t), \pi_{y}(t)) = |\ppi(t)|(\cos \theta_{\kk}(t),\sin \theta_{\kk}(t))$, with: 
\begin{gather}
|\ppi(t)| = \sqrt{\left[ k_{x} + \frac{e}{\hbar c}A(t) \cos \Theta \right]^2 + \left[ k_{y} + \frac{e}{\hbar c} A(t) \sin \Theta \right]^2}   \nonumber \\
\theta_{\kk}(t)  = \arctan \left[ \frac{k_y + e/(\hbar c) A(t) \sin \Theta}{k_x + e/(\hbar c) A(t) \cos \Theta} \right] 
\end{gather}
\nin Extrema of bands of the dispersion of a quasiparticle lead to the identification of \emph{valleys} in momentum space. The goal is to find a suitable Hamiltonian $H^{\xi}_{\kk}(t) $ modelling the dynamics of a carrier of wavevector $\kk$ in a valley $\xi$, within two effective bands. Any $2$-by-$2$  Hermitian operator may be expanded over the Pauli basis $ \{ \sigma_{\nu} ~ | ~ \nu = 0,1,2,3 \}$, where $\sigma_{0}$ is taken as the identity $\mathbb{I}_{2}$ and the remaining respective components as the usual 2D Pauli matrices, i.e. $H_{\kk}(t) = \sum_{\nu} a_{\nu, \kk}(t) \sigma_{\nu}$. For particular real-valued functions $f_{\kk}, r_{\kk}$ and $g_{\kk}$, it can be written in matrix form as:
\begin{equation}
\label{eq:generalhamiltonian}
H^{\xi}_{\kk}(t) = \begin{pmatrix}
f_{\kk}(t)  & e^{-i \xi( \theta_{\kk}(t) - \nu_{\kk}(t))} g_{\kk}(t) \\ 
e^{i \xi( \theta_{\kk}(t) - \nu_{\kk}(t))} g_{\kk}(t) & r_{\kk}(t) 
\end{pmatrix}.
\end{equation}
\nin The newly-introduced phase, termed the \emph{warping phase} $\nu_{\kk}(t)$, appears whenever different hopping contributions (on the off-diagonal entries) rotate at different frequencies. In TMDs, this phase is, perhaps not surprisingly, related to the effect of trigonal warping and this can be seen in Eq.~(\ref{eq:nu_tmd}). \\
\nin Introducing the band index to denote the conduction ($\lambda = +1$) and valence ($\lambda = -1$) bands, the dispersion of each band $\lambda$ is easily obtained through solving $ \text{det}[ H_{\kk}(t) - \epsilon^{\lambda}_{\kk}(t) \mathbb{I}] = 0$. Note that, due to the optical field, this \emph{instantaneous} dispersion is also time-dependent and will be written as:
\begin{equation}
\label{eq:general_dispersion}
 \epsilon^{\lambda}_{\kk}(t) =  \epsilon^{a}_{\kk}(t) + \lambda \epsilon^{b}_{\kk}(t) 
\end{equation}
\nin wherein the contributions to the bands asymmetry are lumped in $\epsilon^{a}_{\kk}$ and the hopping contributions in $\epsilon^{b}_{\kk}$:
\begin{align}
\label{eq:ea_eb}
\epsilon^{a}_{\kk}(t) &= \frac{1}{2} \left( f_{\kk} + r_{\kk} \right) \ & \
\epsilon^{b}_{\kk}(t) &= \sqrt{\left( \frac{f_{\kk} - r_{\kk} }{2}\right)^2 + g^{2}_{\kk}}  
\end{align}
\nin For example, and ignoring the optical coupling, gapless graphene has $\epsilon^{a}_{\kk} = 0$ and $\epsilon^{b}_{\kk} = \hbar v_{\rm F} |\kk|$ --- the usual Dirac cones.\\
\nin The \emph{instantaneous} band eigenstates of the full Hamiltonian solve $H_{\kk}(t) \ket{u^{\lambda}_{\kk}(t)} = \epsilon^{\lambda}_{\kk}(t) \ket{u^{\lambda}_{\kk}(t)}$ and may be written as:
\setlength{\belowdisplayskip}{8pt} \setlength{\belowdisplayshortskip}{0pt}
\setlength{\abovedisplayskip}{5pt} \setlength{\abovedisplayshortskip}{0pt}
\begin{equation}
\label{TMDeigenstates}
\begin{matrix}
\ket{u^{\lambda}_{\kk}} = \left(\frac{g_{\kk}}{\epsilon^{b}_{\kk} \sqrt{2(1-\lambda z_{\kk})}} \right) \begin{pmatrix}
\left( \frac{\eb}{g_{\kk}} \right) (1- \lambda z_{\kk}) e^{-i \xi (\theta_{\kk} - \nu_{\kk})/2} \\ \lambda e^{i \xi (\theta_{\kk} - \nu_{\kk})/2}
\end{pmatrix}
\end{matrix}
\end{equation}
\nin where instantaneous orthonormality $\braket{u^{\lambda}_{\kk}(t)|u^{\lambda'}_{\kk}(t)} = \delta_{\lambda \lambda'}$ was applied. Of further interest, a time-dependent Berry phase $\eta^{\lambda}_{\kk}(t)$ may appear in the dynamics of the carriers; its time derivative is $\dot{\eta}^{\lambda}_{\kk}(t) \equiv i \braket{u^{\lambda}_{\kk}|\dot{u}^{\lambda}_{\kk}} = -\xi \lambda z_{\kk}(\dot{\theta}_{\kk} - \dot{\nu}_{\kk})/2$, where $z_{\kk} \equiv \left( r_{\kk}-\epsilon_{\kk}^{a} \right)/\epsilon_{\kk}^{b}$.
The electric dipole element associated to this two-level system originates from the electron and hole states coupling and is given by: 
\setlength{\belowdisplayskip}{8pt} \setlength{\belowdisplayshortskip}{0pt}
\setlength{\abovedisplayskip}{2pt} \setlength{\abovedisplayshortskip}{0pt}
\begin{equation}
\label{eq:dipolemoment}
 \braket{u^{\lambda}_{\kk}|\dot{u}^{-\lambda}_{\kk}} = 
 -i \xi \frac{g_{\kk}}{2 \epsilon^{b}_{\kk}}(\dot{\theta}_{\kk} - \dot{\nu}_{\kk}) + \lambda \frac{\eb}{2g_{\kk}}\dot{z}_{\kk}  
\end{equation}
\nin The existence of a real part in this element implies that the dipole moment contains an imaginary part. For instance, this feature is discussed and its form shown for massive Dirac fermions in Ref.~\cite{Carvalho2018}, Eq.~(11) . \\

\begin{figure}[ht]
\centering
\includegraphics[scale=0.175]{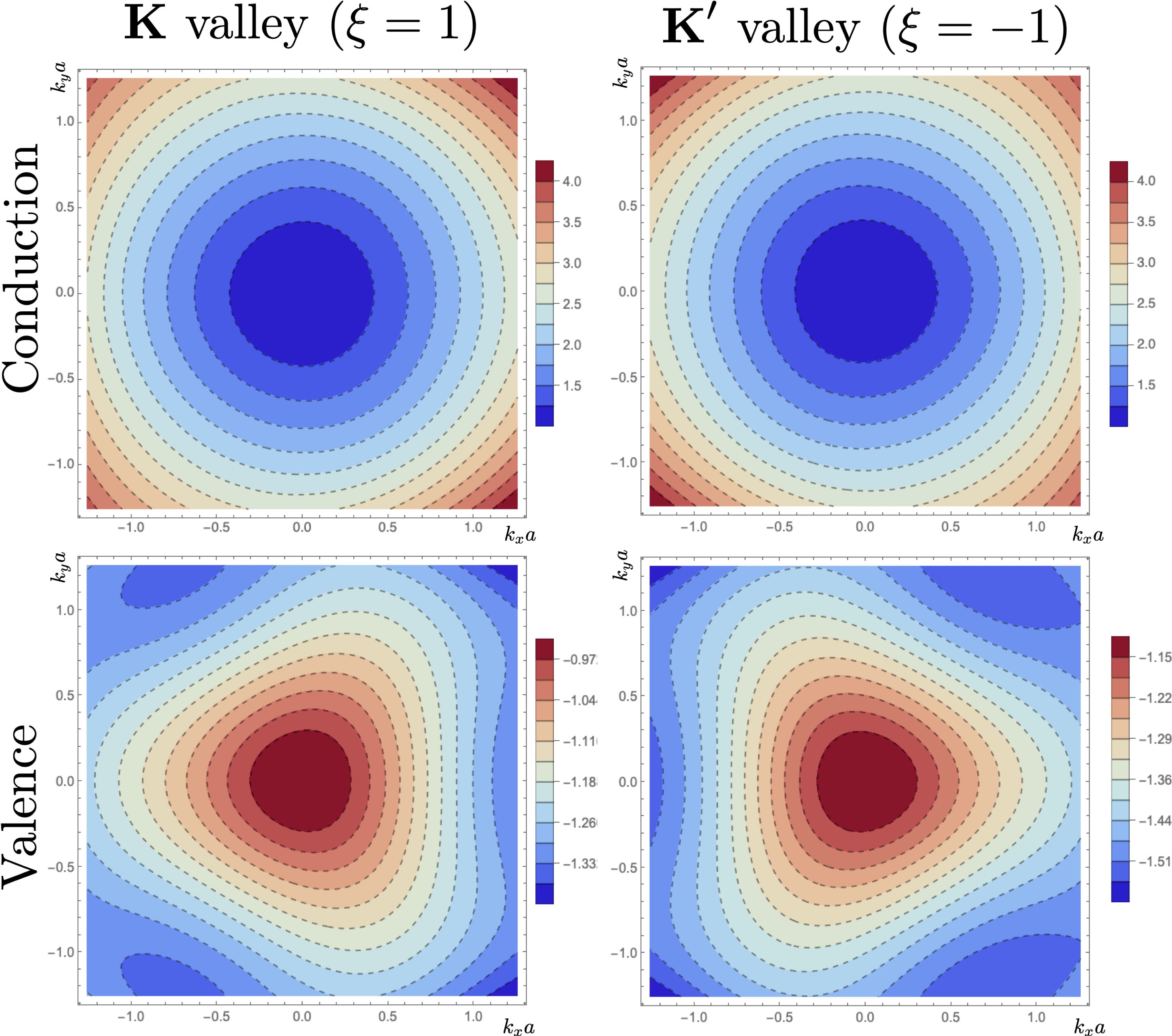}
\caption{Contour plot of the dispersion for the two-band effective model with phenomenological parameters as given by the cubic model in  \ref{sec:cumsim}. The trigonal warping can be seen in contour lines of energy of the valence bands of both valleys.}
\label{fig:contour}
\end{figure}

\nin This procedure so far applies to any system for which an effective two-band model is sensible. In order to obtain the specific functions $a_{\nu, \kk}$ (and consequently $f_{\kk}, r_{\kk}, g_{\kk}, \nu_{\kk}$), which dictate the carrier dynamics across a particular TMD monolayer, approximations are needed. Within the framework of the $\kk \cdot \pp$ expansion \cite{Voon2009}, the Hamiltonian describing the dynamics of the carriers must therefore be expanded up to an adequate order. \\
\nin Several works, e.g., in Refs. \cite{Xiao2012,Yao2014}, only consider a first-order $\kk \cdot \pp$ approximation when modelling the electronics of carriers in TMD monolayers, yielding a massive Dirac fermion description. For low-lying electronic states interacting with optical fields of low intensity, this treatment suffices to capture the linear optical properties of the centrosymmetric medium. However, in order to accurately capture nonlinear light-matter phenomena of noncentrosymmetric media across the entire BZ, such expansion is clearly insufficient. Higher-order terms in the $\kk \cdot \pp$ expansion must be considered so that the centrosymmetry $\kk \leftrightarrow - \kk$ is explicitly broken. This is a key observation and what allows the present treatment to incorporate higher-order electronic contributions, such as trigonal warping, to the nonlinear contributions to the photo-generated current. \\
Furthermore, spin dependences are considered by including a first-order contribution originated from spin-orbit coupling (SOC) effects, leading to a valence band shift. Such a procedure leads to the Hamiltonian written as:
\setlength{\belowdisplayskip}{8pt} \setlength{\belowdisplayshortskip}{0pt}
\setlength{\abovedisplayskip}{8pt} \setlength{\abovedisplayshortskip}{0pt}
\begin{equation}
\label{eq:FullHamiltonian}
H_{\kk}(\xi,s,t) = \sum_{i = 1}^{3}H^{(i)}_{\kk}(\xi,t) + H^{\rm SOC}_{\kk}(\xi,s).
\end{equation} 
\nin Here, $H^{(i)}_{\kk}$ are the i$^{\rm th}$-order corrections in the $\kk \cdot \pp$ expansion, whereas $H^{\rm SOC}_{\kk}$ is the SOC contribution, acting on $s = +1 (-1)~ $ spin up (down) states. Note that in this way, all $4 = 2$ (on $\xi$) $\times~2$ (on $s$) degrees of freedom are incorporated into one expression. For the orbital configurations of TMDs, the exact functions in Eq.~(\ref{eq:generalhamiltonian}) that resulted from the expansion rely on a set of nine phenomenological energy parameters typical of each particular monolayer \cite{Liu2013} -- $\gamma_i ~ (i = 0,...,6)$, $\gamma_{\rm SOC}$ and $\Delta$ -- and may be found in Eq.~(\ref{eq:frgnu}). As for the Hamiltonian operators in Eq.~(\ref{eq:FullHamiltonian}), both their functional form, written as an expansion over the Pauli basis, and their matrix form may be found in the Appendix in Table~\ref{tbl:hamiltonians_TMD}. \\
In order to show the trigonal warping in the valence band and electron-hole symmetry, the dispersion of Eq.~(\ref{eq:general_dispersion}) is plotted in Fig.~\ref{fig:contour}, for both valleys and bands using the parameters of $\tilde{H}^{(3)}_{\kk}$ given in Sec. \ref{sec:cumsim}. The trigonal warping mostly deforms the otherwise-circular Fermi surface on the valence band along the angles $n 2 \pi/3$, $n \in \mathbb{Z}$, leaving the conduction band nearly isotropic. This is supported by \emph{ab-initio} calculations \cite{Zahid2013, Kormanyos2013}, symmetry analysis \cite{Rostami2013} and experimental evidence \cite{Alidoust2014}. Note that the SOC coupling is taken to first-order and therefore not momentum-dependent, leading to a mere shift in the valence bands which cannot be visualised in the figure.

\section{The Two-Band Generalised Dirac-Bloch Equations}
\label{sec:generalised_DBEs}

\nin We seek to obtain the dynamics of a carrier of wavevector $\kk$, spin $s$ in a valley $\xi$, represented by the spinor $\ket{\Psi_{\kk}(\xi,s,t)}$, solution of the Schrödinger Equation:
\setlength{\belowdisplayskip}{5pt} \setlength{\belowdisplayshortskip}{0pt}
\setlength{\abovedisplayskip}{5pt} \setlength{\abovedisplayshortskip}{0pt}
\begin{equation}
\label{SchrodingerEq}
i \hbar \frac{d}{dt} \ket{\Psi_{\kk}(t)} = H_{\kk}(t) \ket{\Psi_{\kk}(t)}
\end{equation}
\nin No analytical solutions can be obtained due to the time-dependence of the Hamiltonian, a consequence of the minimal substitution applied. Despite that, and following methodologies devised in \cite{Ishikawa2010}, an \emph{ansatz} may be taken as a linear superposition of the instantaneous band wavefunctions given in Eq.~(\ref{TMDeigenstates}) like so:
\setlength{\belowdisplayskip}{3pt} \setlength{\belowdisplayshortskip}{0pt}
\setlength{\abovedisplayskip}{8pt} \setlength{\abovedisplayshortskip}{0pt}
\begin{equation}
\label{TMDansatz}
\ket{\Psi_{\kk}(t)} = \sum_{\lambda} c^{\lambda}_{\kk}(t) \ket{u^{\lambda}_{\kk}(t)}e^{i(\eta^{\lambda}_{\kk}(t) - \Omega^{\lambda}_{\kk}(t))} 
\end{equation}
\nin where, for a band $\lambda$, $c_{\kk}^{\lambda}(t)$ are the evolution coefficients, $\Omega^{\lambda}_{\kk}(t)  \equiv (1/\hbar)\int_{-\infty}^{t}\epsilon^{\lambda}_{\kk}(t')dt'$ is the \emph{dynamical phase} and $\eta^{\lambda}_{\kk}(t) = \int_{-\infty}^{t}\dot{\eta}^{\lambda}_{\kk}(t')dt'$ is the \emph{(instantaneous) Berry phase}, whose derivative was given in the previous section. \\
\nin By evaluating the ansatz of Eq.~(\ref{TMDansatz}) in Eq.~(\ref{SchrodingerEq}), the \emph{exact} dynamical evolution of the band coefficients is found. However, it is of interest to transform them to optically-relevant variables for a two-level system: the \emph{inversion} $w_{\kk}(t) \equiv |c^{(+)}_{\kk}|^2-|c^{(-)}_{\kk}|^2$ and the \emph{microscopic polarisation} (or \emph{coherence}) $q_{\kk} \equiv c^{(+)}_{\kk} {c^{(-)}_{\kk}}^{*} \exp(i(\omega_{0}t - 2 \Omega_{\kk}^{b}))$, where $\omega_{0}$ is the pulse central frequency and $\Omega^{b}_{\kk}(t)  \equiv (1/\hbar)\int_{-\infty}^{t}\epsilon^{b}_{\kk}(t')dt'$. The dynamics of these fields is described by the \emph{Two-Band Generalised Bloch Equations}: 
\setlength{\belowdisplayskip}{8pt} \setlength{\belowdisplayshortskip}{0pt}
\setlength{\abovedisplayskip}{8pt} \setlength{\abovedisplayshortskip}{0pt}
\begin{equation}
\label{eq:dbes}
\begin{matrix}
\dot{q}_{\kk} - i\left(\omega_{0} - 2 \dot{\Omega}^{b}_{\kk} \right) q_{\kk} - \braket{u^{(+)}_{\kk}|\dot{u}^{(-)}_{\kk}}e^{i(\omega_{0}t + \xi \Lambda_{\kk})} w_{\kk} =0\\
\dot{w}_{\kk} + 4 \text{Re}\left(
\braket{u^{(+)}_{\kk}|\dot{u}^{(-)}_{\kk}}^{*} e^{-i(\omega_{0}t + \xi \Lambda_{\kk})} q_{\kk} \right) = 0
\end{matrix}.
\end{equation}
\nin with $\Lambda_{\kk}(t) \equiv - 2 \xi \lambda  \eta^{\lambda}_{\kk}(t)$ band-independent. Since no dephasing mechanisms were included, the quantity $4 |q_{\kk}|^2 +  w_{\kk}^2$ is conserved. Otherwise, phenomenological decay rates may be included.
\nin Note that only the carrier-free contributions have been considered in the dynamics. Many-body correlations, including non-negligible electron-electron interactions are not accounted here.

\subsection{Photo-generated current}
\label{sec:current}

\begin{figure}[ht]
\label{fig:kp_expansion_current}
\centering
\includegraphics[scale=0.20]{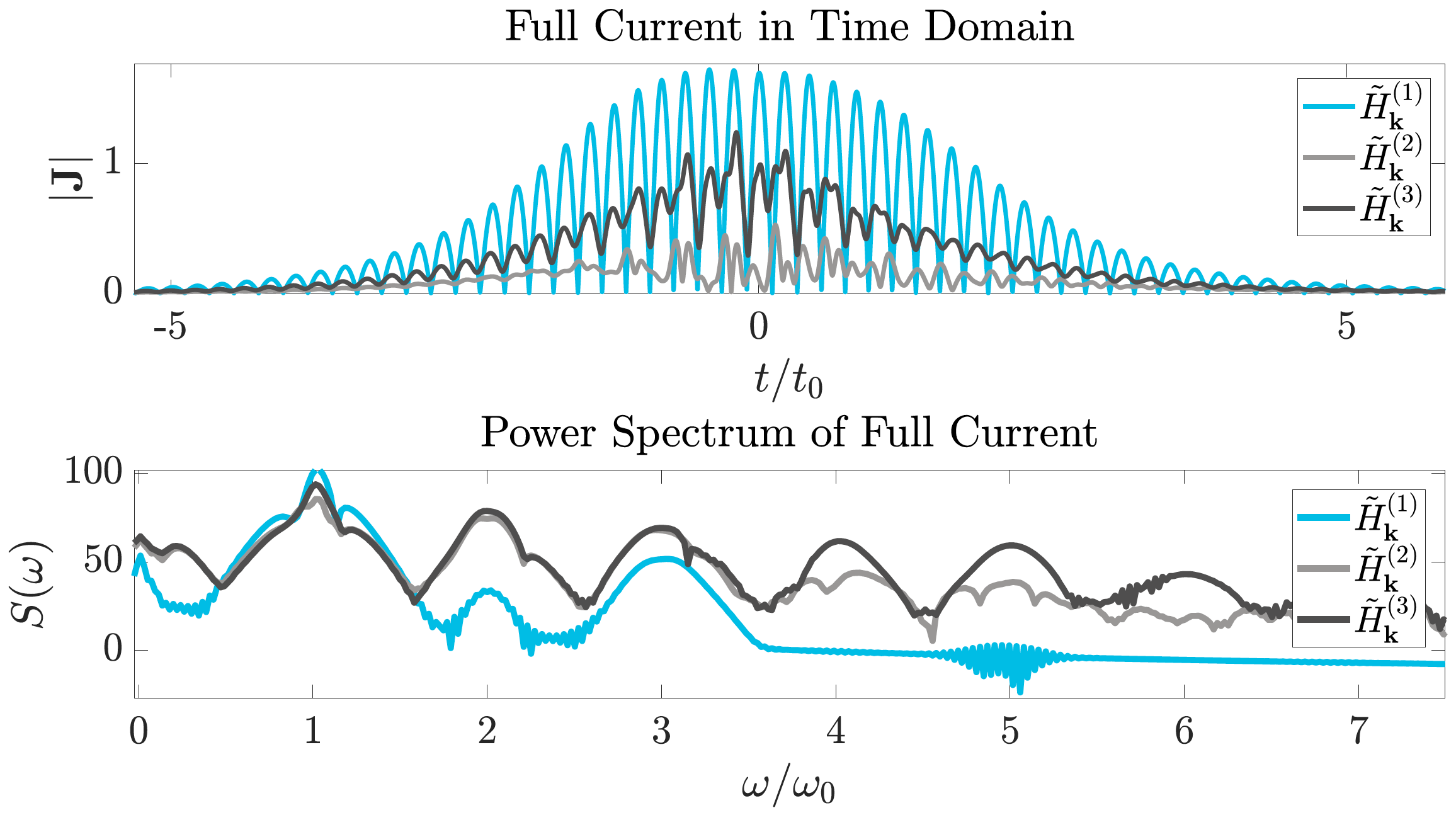}
\caption{Full output current in (a) time domain and (b) frequency domain. The addition of $\kk \cdot \pp$ terms leads to strong frequency and amplitude modulation of the current when the optical field is maximal, at $t/t_0 = 0$. This is confirmed in its respective spectrum, where strong even harmonics are created with the addition of explicit centrosymmetry-breaking terms in the quadratic and cubic models.}
\end{figure}

\nin With the knowledge of how to obtain the dynamics of the carriers when coupled to the optical field, we now seek to obtain the surface current $\mathbf{J}(t) = (J_{x}(t), J_{y}(t))$ generated across the sample as a result of such light-matter interactions. To this end, the \emph{microscopic} contributions of each $\kk$ state are obtained first. Due to the $\kk \cdot \pp$ expansion applied in Eq.~(\ref{eq:FullHamiltonian}), and noting that $H^{\rm SOC}_{\kk}$ does not contribute to the current as it is not $\kk$ dependent, the operator associated to the current density in a Cartesian coordinate $\mu \in \{ x,y \}$ may be written as $\hat{j}_{\mu, \kk}(t) = \sum_{i} \hat{j}^{(i)}_{\mu, \kk}(t)$. A natural definition of each contribution is the so-called paramagnetic current operator
\begin{equation}
\label{eq:currentoperator}
\hat{j}^{(i)}_{\mu, \kk}(t) \equiv - c \frac{\delta}{\delta A_{\mu}}\left( H^{(i)}_{\kk}(t) \right) \vert_{A_{\mu} \rightarrow 0}
\end{equation}
\nin For the wavefunction in Eq.~(\ref{TMDansatz}), each operator leads to the current:
\setlength{\belowdisplayskip}{10pt} \setlength{\belowdisplayshortskip}{0pt}
\setlength{\abovedisplayskip}{10pt} \setlength{\abovedisplayshortskip}{0pt}
\begin{equation}
\label{eq:current_observable}
j^{(i)}_{\mu, \kk}(t) =\braket{\Psi_{\kk}|\hat{j}^{(i)}_{\mu, \kk}|\Psi_{\kk}} - \braket{u^{(-)}_{\kk}|\hat{j}^{(i)}_{\mu, \kk}|u^{(-)}_{\kk}}
\end{equation}
\nin The first term contains \emph{intraband} and \emph{interband} contributions, depending on whether currents originate from the same or opposite bands, respectively. The latter term must be considered in order to regularise the otherwise divergent first term. If the regularisation term is incorporated in the intraband current, the current may be written as:
\begin{equation}
j^{(i)}_{\mu, \kk}(t) = j^{\rm intra (i)}_{\mu, \kk}(t) + j^{\rm inter (i)}_{\mu, \kk}(t)
\end{equation}
\nin In this fashion, the current generation is predicted \emph{exactly} and \emph{nonperturbatively} by knowing the two-level system dynamics through the DBEs of Eq.~(\ref{eq:dbes}) as:
\setlength{\belowdisplayskip}{10pt} \setlength{\belowdisplayshortskip}{0pt}
\setlength{\abovedisplayskip}{5pt} \setlength{\abovedisplayshortskip}{0pt}
{\renewcommand{\arraystretch}{1.75}
\begin{equation}
\label{eq:intraintercurrents}
\begin{matrix}
j^{\rm intra (i)}_{\mu, \kk} = \sum_{\nu = 1}^{3} a^{(i)}_{\mu, \nu}(\kk) \braket{u^{(+)}_{\kk}|\sigma_{\nu}|u^{(+)}_{\kk}} (w_{\kk} + 1) \\
j^{\rm inter (i)}_{\mu, \kk} = \sum_{\nu} 2 a^{(i)}_{\mu, \nu}(\kk) \text{Re} \left( q_{\kk} e^{-i( \xi \Lambda_{\kk} + \omega_{0}t)} \braket{u^{(-)}_{\kk}|\sigma_{\nu}|u^{(+)}_{\kk}} \right) 
\end{matrix}
\end{equation}
\nin For the case of Mo$S_2$ monolayers, the exact functional form of these currents for all orders and Cartesian components is rather cumbersome and may be found in Table~\ref{tbl:intraband_currents_TMD} (intraband) and Table~\ref{tbl:interband_currents_TMD} (interband). \\
\nin Consequently, the current may be estimated once the system in study is well resolved in momentum space i.e. when $a_{\mu, \nu}^{(i)}(\kk)$ is known. One can see in Eq.~(\ref{eq:intraintercurrents}) that the intraband currents depend on the inversion $w_{\kk}$, viz the carrier populations while the interband contributions are dictated by the polarisation $q_{\kk}$. Direct calculation of these terms shows that two quantities determine the current composition for either contribution; the ratio $g_{\kk}/\epsilon_{\kk}^{b} ~ (z_{\kk})$ expresses the relative contribution of the hopping (on-site) terms which appear as off-diagonal (diagonal) terms in the Hamiltonian of Eq.~(\ref{eq:generalhamiltonian}). These strengths are in fact constrained through $(g_{\kk}/\epsilon_{\kk}^{b})^2 + z_{\kk}^2 = 1$. \\
\nin Note that each contribution $j_{\mu, \kk}^{(i)}$ should contain in principle all harmonic contributions due to light i.e. only the electronic field is being expanded and not the optical field. Finally, the physical current is retrieved by averaging all microscopic contributions of both valleys and spin states into account. In the continuum limit, it is:
 \begin{equation}
\label{macrocurrent}
\mathbf{J}(t) = \frac{1}{d(2 \pi)^2} \sum_{\xi,s} {\int \mathbf{j}_{\kk}(\xi,s,t)}d \kk,
\end{equation}
\nin where $d$ is the thickness of the monolayer, $d \kk = k dk d \phi$ is the 2-dimensional differential in momentum space. In order to obtain information in the frequency domain, the power spectrum $S(\omega)$ (in dBs) is introduced as:
\begin{equation}
\label{eq:currentspectrum}
S(\omega) = 10 \log_{10} \left(\omega^2 |\tilde{\JJ}(\omega)|^2 \right)
\end{equation}
\nin where $\tilde{\JJ}(\omega)$ is the Fourier transform of $\JJ(t)$. 

\section{Simulations}
\label{sec:simulations}

\nin Given the suitability of the Hamiltonian expansion given in Table~\ref{tbl:hamiltonians_TMD} to fit ab-initio band calculations of MoS$_2$ (as may be appreciated in Fig. 10 of Ref. \cite{Liu2013}) within a reasonable vicinity range of the valleys, and the apparent optoelectronic superiority in comparison to other TMDs \cite{Singh2018, Marini2018}, we use this system as an illustrative example of how to study nonlinear signatures using the machinery so far introduced. However, we emphasise that the model introduced in this article is applicable for single-particle effective two-level systems for which the tight-binding parameters are known, without any need for low-energy $\kk \cdot \pp$  expansion. \\

\nin For any analysis of the current to be performed, the DBEs [Eqs.~(\ref{eq:dbes})] must be solved numerically. We choose a normally-incident pulse of duration $t_0 = 31.9$ fs, central frequency  $\omega_0 = 4.71 \times 10^{-14}$ s$^{-1}$, photon energy $\hbar \omega_0 = 0.31$ eV, intensity $I = 0.45$ GW/cm$^{2}$ with vector potential $A(t) = A_0 \sech{t/t_0} \sin (\omega_0 t)$ and electric field $E(t) = -(1/c) \partial A/ \partial t$. Furthermore, we restrict our attention to samples at temperature $T = 0 ^{\circ}$ K, admitting carriers with a vanishing Fermi level and perfectly coherent i.e. $\gamma_1 = \gamma_2 = 0$.

\subsection{Role of the $\kk \cdot \pp$ expansion}
\label{sec:cumsim}

\nin In order to see how the formal introduction of further terms in the $\kk \cdot \pp$ expansion impact the harmonic generation predicted by the model, we proceed by defining the truncated Hamiltonians containing the cumulative contributions $\tilde{H}_{\kk}^{(i)} \equiv \sum_{j=1}^{i} H_{\kk}^{(j)} + H_{\kk}^{\rm SOC}$. 
\nin The phenomenological parameters are taken from \cite{Liu2013} where the lattice constant is $a = 3.19$ \AA ~ and the gap is $\Delta = 1.663$ eV; the remaining constants are denoted here by $\mathbf{g} \equiv (g_{i} ~|~ i = 0, ~..., ~ 6 )$ and expressed in eV. Physically speaking, $g_0$ sets the Fermi velocity $v_{\rm F} \equiv a \gamma_0/\hbar$,$\gamma_1,2,4,5,6$ quantity electron and hole asymmetries in second and third order and $\gamma_3$ sets the strength of trigonal warping. \\
\nin The linear mode $\tilde{H}^{(1)}_{\kk}$ has vanishing entries apart from $\gamma_0 = 1.105$ and it may be seen as a suitable electronic model of doped graphene. The quadratic model $\tilde{H}^{(2)}_{\kk}$  has ${\bf g} = (1.059, 0.055, 0.077, -0.123, 0,0,0 )$ and $\gamma_{\rm SOC} = 0$. The cubic model $\tilde{H}^{(3)}_{\kk}$  has ${\bf g} = (1.003, 0.196, -0.065,  -0.248, 0.163 , -0.094, -0.232 ) $ and $\gamma_{\rm SOC} = 89.6$ meV. Note that each model uses different corresponding parameters. \\

\nin Due to the linearity of the derivatives, the corrections to the current may be obtained individually by computing the operators in Eq.~(\ref{eq:currentoperator}) and applying them to the general wavefunction as prescribed in Eq.~(\ref{eq:current_observable}). Therefore, the explicit effect of the anisotropy may be explicitly inspected in the harmonic generation. \\ In Fig. \ref{fig:kp_expansion_current}}, the absolute value of the full current defined in Eq.~(\ref{macrocurrent}), whose integrand is explicitly given in Table~\ref{tbl:intraband_currents_TMD} - \ref{tbl:interband_currents_TMD}, is shown in  Fig.~\ref{fig:kp_expansion_current}(a) for the three $\kk \cdot \pp$ Hamiltonians $\tilde{H}^{(i)}_{\kk},~ i = 1,2,3$. It can be seen that the current roughly follows the pulse profile in time domain. However, both the frequency and amplitude modulation of the signal deviate from each other once the optical field is maximal (the pulse peak is centred about $t=0$) as higher $\kk \cdot \pp$ terms are included. We shall focus on the total, physically-relevant current, which is composed of both intraband and interband contributions. However, we remark that, given the resonant pumping conditions, the interband contributions dominate over 
their intraband counterparts, although the intraband current amplitude increases for higher $\kk \cdot \pp$ models, contrary to the amplitude suppression shown in Fig. \ref{fig:kp_expansion_current}(a).  \\

\nin The effects of this modulation may be seen in the harmonic composition of the currents in Fig.~\ref{fig:kp_expansion_current}(b), where the power spectrum of Eq.~(\ref{eq:currentspectrum}) is plotted in units of the pulse-normalised frequency $\omega/\omega_0$. The linear model shows the well-known strong third harmonic observed in graphene, as well as small second-harmonic peaks previously reported to exist due to light-induced dynamical centrosymmetry breaking \cite{Carvalho2017, Carvalho2018}. 
As expected, the explicit centrosymmetry breaking of the lattice arrangement as depicted in Fig.~\ref{fig1} induces strong even harmonics, which are observed once quadratic and cubic terms are added. Linked to second harmonic generation, optical rectification can also be seen to increase in the same fashion in the DC limit (at $\omega = 0$). This turns out to be an interband-driven process.

\nin We close this subsection with a due remark. The output current $\JJ$ has, perhaps surprisingly, both non-vanishing longitudinal and transverse components to the electric field, polarised  along $\Theta$. The present machinery has been previously applied to both gapless and gapped graphene monolayers in \cite{Carvalho2017,Carvalho2018}. The gapless case presented a straightforward analysis: only current along the same direction as the electric field was found upon integration in Eq.~(\ref{macrocurrent}), where a degeneracy factor $g_v = 2$ was verified. Once the system became gapped, the transverse current still vanished, but now due to a cancellation of valley transverse currents. With the introduction of further $\kk \cdot \pp$ terms, this no longer holds. To verify the invariance of this result, we applied a rotation $\mathcal{R}(\alpha)$ to the basis $\{\hat{x},\hat{y}\}$ and checked if $\mathcal{R}(\alpha)\JJ(t)$ admits only one non-vanishing component. No angle was found to yield this result, leading us to conclude these corrections to the dispersion may indeed create transverse currents.

\subsection{Role of the electromagnetic polarisation}
\label{sec:directionalsim}

\begin{figure}[ht]
\centering
\includegraphics[scale=0.20]{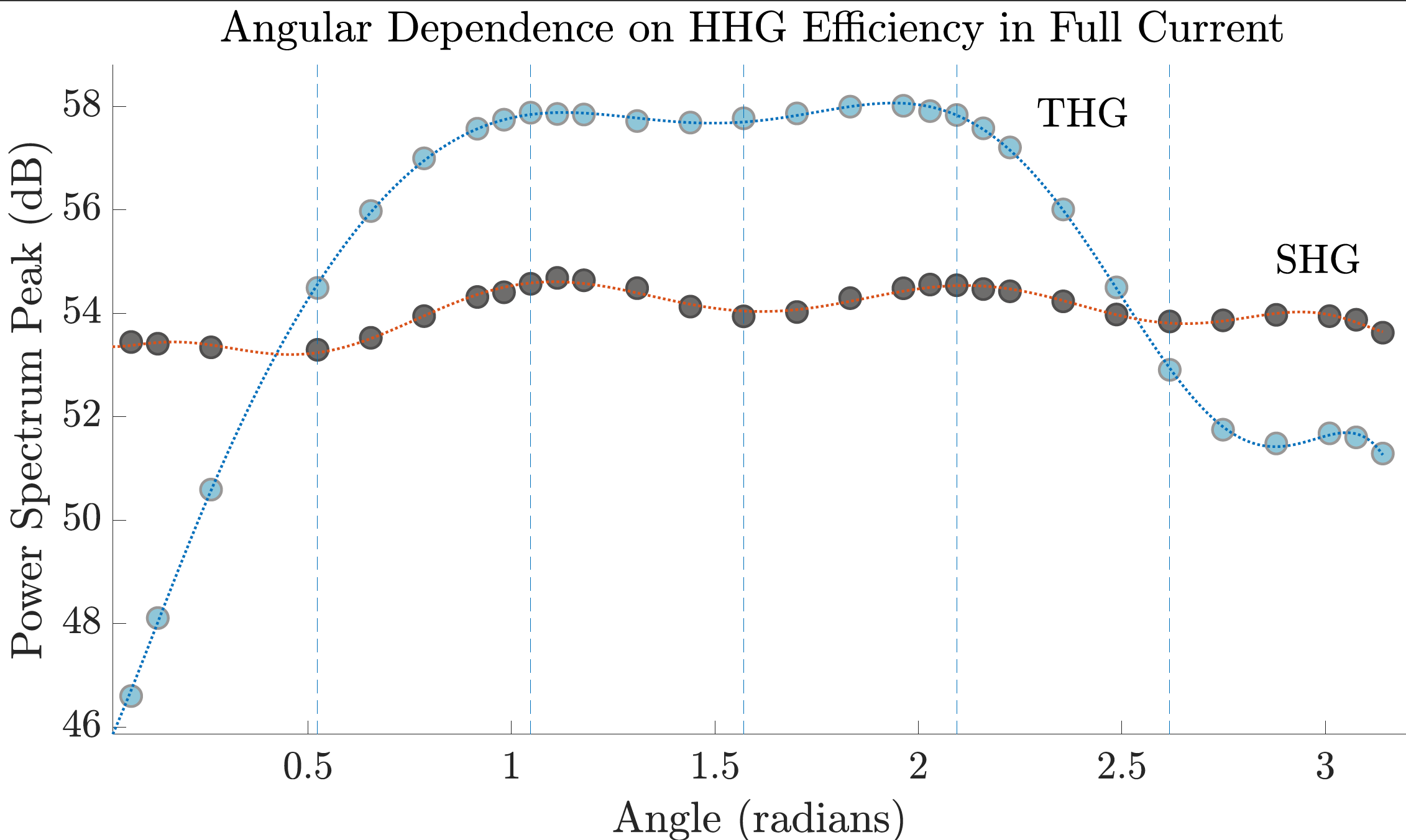}
\caption{Maxima of the second harmonic (light gray) and third harmonic (dark gray) as a function of the polarisation angle $\Theta$ for the same pumping conditions. The modulation of the second harmonic is seen to have the same period of the trigonal warping., with the aid of fitting polynomials. The dashed lines correspond to multiples of $\pi/6$. Within the range $\pi/6 \leq \Theta_{\rm c} \leq 5 \pi/6$, the THG overcomes the SHG intensity by 4 orders of magnitude.}
\label{fig5}
\end{figure}

\nin Given the anisotropy properties of the dispersion shown in shown in Fig.\ref{fig1}, it is expected that the photo-generated current across the sample depends on the relative angle between the lattice orientation and the oscillation of the electric field. This the polarisation angle $\Theta$ is measured on the sample plane and depicted in Fig.\ref{fig:light}. In order to appreciate such dependence, a MoS$_2$ monolayer is excited using the phenomenological energy parameters of $\tilde{H}^{(3)}_{\kk}$ in resonance conditions i.e. with $\Delta = 2 \hbar \omega_0$, when sweeping through the range $0 \leq \Theta \leq \pi$. For each angle, the momentum-integrated spectrum of Eq.~(\ref{eq:currentspectrum}) is calculated and the maximum intensity of each harmonic peak is recorded. Such plot is shown in Fig.\ref{fig5} for 
second (SHG) and third harmonic (THG)-generated peaks. To help visualise the trend, a polynomial fit is added to each harmonic. \\

\nin To relate the anisotropy to the polarisation modulation, dashed lines of strong trigonal warping are added increasingly at $\Theta = \pi/3, \pi/6, \pi/2, 2 \pi/3$ and $5 \pi/6$. The SHG signal shows a consistent modulation, maxima at $\Theta \approx n \pi/3$ and minima at $\Theta \approx (2n + 1) \pi/6$, leading to the same period of the trigonal warping, of $\pi/3$. The THG modulation shows somewhat similar patterns as SHG but differs at small angles, showing a very large variation (note the scale is logarithmic). Higher harmonic peaks show more complex, nontrivial modulation. Their exact functional form is not instructive for our study since, as the harmonic order increases, the efficiency of nonlinear optical processes decays exponentially and are therefore not shown. \\
\nin Here, we focus on the possibility of harmonic crossovers i.e. higher, less efficient harmonics amplitudes to overcome lower, more efficient harmonics. Since harmonics of all parity may be created in the present cubic model, as demonstrated in Fig.~\ref{fig:kp_expansion_current}(b), this possibility is tested. Indeed, a crossover range where the third-harmonic is stronger than the second-harmonic is found, namely between $\pi/6 \leq \Theta_{\rm c} \leq 5 \pi/6$. This difference is about 4 orders of magnitude, meaning that it could potentially be measured in experimental conditions.

\section{Conclusions}
\label{sec:conclusions}

\nin In conclusion, we have generalised the Dirac-Bloch  Equations (DBEs) and their formalism to any single quasiparticle described by an effective two-band model. Through the temporal evolution of the carriers, as dictated by the Dirac equation, the current generated by an incoming linearly-polarised pulse was obtained in a non-perturbative fashion. Intraband and interband current contributions may be computed once the momentum-dependent hoppings and on-site energies are known. \\
\nin As a prototypical example, we study higher-order corrections in $\kk \cdot \pp$ in the nonlinear response of MoS$_2$ monolayers, representative of a wider range of promising new semiconducting materials known as transition metal dichalcogenides. Owing to their orbital character, the Fermi surface of allowed quasiparticle excitations becomes warped, leading to a three-fold anisotropy in the dispersion. \\

\nin  We studied the effect of such a feature on the photo-generated nonlinear current by applying an effective tight-binding model, expanded up to third-order in $\kk \cdot \pp$. The anisotropy seems to lead to transverse currents with respect to the electric field excitation and non-trivial frequency and amplitude modulation of the current in time domain. We observe interband-driven enhancement of even harmonic generation, as well as intraband-driven enhancement of optical rectification as quadratic and cubic terms are added, explicitly breaking the centrosymmetry of the crystal. \\
\nin To capture anisotropy-induced optical effects, we compute the current for various polarisation angles of the pulse with respect to the lattice and conclude that the harmonic peak intensity is modulated periodically by the warping profile. For a tuning region of values, third harmonic generation was shown to dominate over second harmonic generation by four orders of magnitude, suggesting this effect could be realised experimentally. \\

\nin The machinery introduced may well be used to model many other condensed matter systems. For instance, effective two-band models have been proposed for bilayer graphene \cite{Mccann2013} and topological insulators \cite{Liu2010}.\\
\nin Of great conceptual importance in this work, we remark that the light contribution to the dynamics is not taken perturbatively, implying that the output current should contain contributions of all orders in the electric field. \\
\nin Despite this advantage for studying nonlinear optics, this methodology does not take into account Coulomb interactions of the carriers and therefore still admittedly incomplete and incapable of capturing other relevant physics. Further generalisations are possible; for instance, the Coulomb interactions have been included in the the Dirac-Bloch modelling graphene \cite{DiMauro2018}. \\
\nin This work hopes to inspire new methodologies to study features of anisotropic media and protocols to fine-tune harmonic generation in heterostructures and other nanophotonic devices.

\begin{acknowledgements}

D.N.C. wishes to thank Leone di Mauro Villari for insightful discussions and computational support. This work has been supported by the International Max Planck Partnership (IMPP) between Heriot-Watt University and the Max-Planck Society and the Scottish Universities Physics Alliance (SUPA).

\end{acknowledgements}

\onecolumngrid
\appendix

\section{Hamiltonian Parameters}

\nin For MoS$_2$, the energy parameters of Eq.~\ref{eq:generalhamiltonian} take the functional form \cite{Liu2013}:

\begin{subequations}
\label{eq:frgnu}
\begin{gather}
f_{\kk} = \frac{\Delta}{2} +  \gamma_{1} a^2|\ppi|^2 + \xi \gamma_{4}  \cos(3 \theta_{\kk}) a^{3} |\ppi|^3    \\
r_{\kk} = -\frac{\Delta}{2}  + s \xi \gamma_{\rm SOC} +  \gamma_{2} a^2 |\ppi|^2 + \xi \gamma_{5} \cos(3 \theta_{\kk}) a^{3} |\ppi|^3 \\
\nu_{\kk} = \arctan \left[\frac{ \gamma_{3} \sin (3 \theta_{\kk}) a |\ppi|}{\xi \gamma_{0}  + \gamma_{3}   \cos(3 \theta_{\kk}) a |\ppi| + \xi  \gamma_{6} a^2 |\ppi|^2} \right] \label{eq:nu_tmd} \\
g_{\kk} = \sqrt{ \gamma^2_{0} a^2|\ppi|^2 + 2 \xi  \gamma_{0} \gamma_{3} \cos (3 \theta_{\kk}) a^3 |\ppi|^3 + (\gamma^{2}_{3} + 
2 \gamma_{0} \gamma_{6})
a^4 |\ppi|^4 \nonumber  \ + 2 \xi \gamma_{3} \gamma_{6} \cos(3 \theta_{\kk})a^5 |\ppi|^5 +  \gamma^{2}_{6} a^{6}|\ppi|^6} \\
\end{gather}
\end{subequations}
\nin With $\ssigma(\xi) = (\xi \sigma_{x}, \sigma_{y})$, each Hamiltonian contribution in the $\kk \cdot \pp$ expansion admits the following operators, which can be easily expanded in the Pauli basis $\{ \sigma_{\nu} \}$ in order to extract the coefficients $a_{\nu, \kk}$.

{\renewcommand{\arraystretch}{2.5}
\begin{table*}[h]
\begin{ruledtabular}
\begin{tabular}{c|c|l}
\textbf{Hamiltonian} & \textbf{Operator Form}
 & \ \ \ \ \ \ \ \ \ \textbf{Matrix Form} \\
 \hline
$H^{(1)}_{\kk}$ & $a \gamma_0 \ssigma(\xi) \cdot \ppi + \frac{\Delta}{2} \sigma_{z}$ & $
\begin{aligned} \ \ \ \ \ \ \ \ \ \ 
\begin{pmatrix}
\frac{\Delta}{2} & \xi a \gamma_0 |\ppi|e^{-i \xi \theta_{\kk}} \\ 
\xi a \gamma_0 |\ppi|e^{i \xi \theta_{\kk}} & -\frac{\Delta}{2}
\end{pmatrix}
\end{aligned}$ \\
$H^{(2)}_{\kk}$ & $a^2 \left( \gamma_{3}(\ssigma(\xi) \cdot \ppi)^{*}\sigma_{x} (\ssigma(\xi) \cdot \ppi)^{*} + \frac{|\ppi|^2}{2} \left[(\gamma_{1} + \gamma_{2}) \mathbb{I} + (\gamma_{1} - \gamma_{2}) \sigma_{z} \right] \right)$ & $
\begin{aligned}
\ a^2 |\ppi|^2 & \begin{pmatrix}
\gamma_{1} & \gamma_{3} e^{2 i \xi \theta_{\kk}} \\ 
\gamma_{3} e^{-2 i \xi \theta_{\kk}} & \gamma_{2}
\end{pmatrix}
\end{aligned}$ \\ 
$H^{(3)}_{\kk}$ & $a^3 \left(\gamma_{6} |\ppi|^2 \ssigma(\xi) \cdot \ppi +\frac{\xi}{2} |\ppi|^3 \cos (3 \theta_{\kk}) \left[(\gamma_{4} + \gamma_{5}) \mathbb{I} + (\gamma_{4} - \gamma_{5}) \sigma_{z} \right] \right)$ & $
\begin{aligned}	
\xi a^3 |\ppi|^3 & \begin{pmatrix}
\gamma_{4} \cos(3 \theta_{\kk}) & \gamma_{6}e^{-i \xi \theta_{\kk}} \\ 
\gamma_{6}e^{i \xi \theta_{\kk}} & \gamma_{5} \cos(3 \theta_{\kk})
\end{pmatrix}
\end{aligned}
$ \\
 $H^{\rm SOC}_{\kk}$ & $\frac{\xi s \gamma_{\rm SOC}}{2} \left[\mathbb{I} - \sigma_{z} \right]$ & $\begin{aligned} \ \ \ \ \ \ \ \ \ \ \ 
 & \begin{pmatrix}
0 & 0 \\ 
0 & \xi s \gamma_{\rm SOC} \end{pmatrix}
\end{aligned}$
\end{tabular}
\end{ruledtabular}
\caption{Terms of the $\kk \cdot \pp$ expansion of the full Hamiltonian, up to third order. $a$ denotes the lattice constant of the TMD monolayer.}
\label{tbl:hamiltonians_TMD}
\end{table*} 

\section{Nonperturbative photo-generated current in MoS$_2$/TMD}

{\renewcommand{\arraystretch}{2.3}
\begin{table*}[]
\begin{ruledtabular}
\begin{tabular}{cc|c}
\multicolumn{1}{c}{} & \textbf{Order} & 
\textbf{Microscopic intraband current  contributions} $j^{\rm intra (i)}_{x, \kk}(\xi, s, t)$ \ \ \ \ \ \ \ \ \ \ \ \ \ \ \ \ \ \ \ \  \\
 \hline
 & $1^{\rm st}$     &       
 $ \begin{aligned}
 j^{\rm intra(1)}_{x, \kk} = -\frac{\xi e a \gamma_{0} g_{\kk}}{\hbar \epsilon^{b}_{\kk}} \cos(\theta_{\kk} - \nu_{\kk}) (w_{\kk}+1)
 \end{aligned}$ \\
  
& $2^{\rm nd}$     &      
$ \begin{aligned}
j^{\rm intra(2)}_{x, \kk} = -\frac{e a^2 |\ppi|}{\hbar} \left( \frac{2 \gamma_{3}g_{\kk}}{\epsilon^{b}_{\kk}} \cos(2 \theta_{\kk} - \nu_{\kk}) - (\gamma_{1} - \gamma_{2})z_{\kk} \cos \theta_{\kk} \right) (w_{\kk} + 1)
  \end{aligned}
$ \\
\multirow{3}{*}{\rot{\rlap{\textbf{ \ \ \ \ $x$ component}}}} 
 & $3^{\rm rd}$     &     
$\begin{aligned}
j^{\rm intra(3)}_{x, \kk} = -\frac{\xi e a^3 |\ppi|^2}{\hbar}  \Biggl( \frac{\gamma_{6} g_{\kk}}{\epsilon^{b}_{\kk}} \left( 3 \cos \theta_{\kk} \cos \nu_{\kk} + \sin \theta_{\kk} \sin \nu_{\kk}\right) - \frac{3 (\gamma_{4} - \gamma_{5})}{2} z_{\kk} \cos 2 \theta_{\kk}  \Biggr) (w_{\kk} + 1) 
\end{aligned}$ \\
 \hline          
           
  & $1^{\rm st}$     &       $ \begin{aligned}
j^{\rm intra(1)}_{y, \kk} =  -\frac{\xi e a \gamma_{0} g_{\kk}}{\hbar \epsilon^{b}_{\kk}} \sin(\theta_{\kk} - \nu_{\kk}) (w_{\kk}+1)
\end{aligned}$ \\
 & $2^{\rm nd}$     &      $\begin{aligned}
j^{\rm intra(2)}_{y, \kk} = -\frac{e a^2 |\ppi|}{\hbar} \left( -\frac{2 \gamma_{3}g_{\kk}}{\epsilon^{b}_{\kk}} \sin(2 \theta_{\kk} - \nu_{\kk}) - (\gamma_{1} - \gamma_{2})z_{\kk} \sin \theta_{\kk} \right) (w_{\kk} + 1) 
\end{aligned}
$ \\
\multirow{3}{*}{\rot{\rlap{\ \ \ \ \ \ \textbf{$y$ component}}}} & $3^{\rm rd}$     &     
$\begin{aligned}
j^{\rm intra(3)}_{y, \kk} = -\frac{\xi e a^3 |\ppi|^2}{\hbar}  \Biggl( \frac{\gamma_{6} g_{\kk}}{\epsilon^{b}_{\kk}} \left( 2 \sin (\theta_{\kk} - \nu_{\kk}) + \sin(\theta_{\kk} + \nu_{\kk})\right)  + \frac{3(\gamma_{4} - \gamma_{5})}{2} z_{\kk} \sin 2 \theta_{\kk}  \Biggr) (w_{\kk} + 1)
\end{aligned}$                          
\end{tabular}
\end{ruledtabular}
\caption{\label{tbl:intraband_currents_TMD} Contributions to the microscopic intraband current up to third $\kk \cdot \pp$ order, for both Cartesian components.}
\end{table*}
{\renewcommand{\arraystretch}{3}
\begin{table*}[]
\begin{ruledtabular}
\begin{tabular}{cc|r}
\multicolumn{1}{c}{} & \textbf{Order} & 
\textbf{Microscopic interband current  contributions} $j^{\rm inter (i)}_{x, \kk}(\xi, s, t)$ \ \ \ \ \ \ \ \ \ \ \ \ \ \ \ \ \ \ \ \  \\
 \hline
 & $1^{\rm st}$     &       $ \begin{aligned}
j^{\rm inter (1)}_{x, \kk} = -\frac{2 \xi e \gamma_{0} a}{\hbar} \Biggl( z_{\kk} \cos(\theta_{\kk} - \nu_{\kk}) \text{Re}\left(q_{\kk} e^{-i(\xi \Lambda_{\kk} + \omega_{0}t)} \right) - \xi \sin \theta_{\kk} \text{Im} \left( q_{\kk} e^{-i(\xi \Lambda_{\kk} + \omega_{0}t)} \right) \Biggr)
\end{aligned}$ \\ 
\multirow{3}{*}{\rot{\rlap{\textbf{$x$ component}}}} 
& $2^{\rm nd}$     &      $\begin{aligned}
j^{\rm inter (2)}_{x, \kk} = -\frac{2 e a^2 \ppi}{\hbar}\Biggl( \Biggl( 2 \gamma_{3}  z_{\kk}\cos(2 \theta_{\kk} - \nu_{\kk}) & + (\gamma_{1}-\gamma_{2})\frac{g_{\kk}}{\epsilon^{b}_{\kk}} \cos \theta_{\kk}  \Biggr)
\text{Re}\left( q_{\kk} e^{-i(\xi \Lambda_{\kk} + \omega_{0}t)}\right) \\ & - 2 \gamma_{3} \xi \sin(2 \theta_{\kk}-\nu_{\kk}) \text{Im}\left( q_{\kk} e^{-i(\xi \Lambda_{\kk} + \omega_{0}t)}\right) \Biggr)
\end{aligned}
$ \\
 & $3^{\rm rd}$     &     
$\begin{aligned}
j^{\rm inter (3)}_{x, \kk} = -\frac{e a^3 |\ppi|^2}{\hbar} \Biggl( \Bigl( 2 \xi \gamma_{6} z_{\kk}( 3 \cos \theta_{\kk} \cos \nu_{\kk} + & 
\sin  \theta_{\kk} \sin \nu_{\kk}) 
 + 3 \xi (\gamma_{4} - \gamma_{5}) \frac{g_{\kk}}{\epsilon^{b}_{\kk}} \cos 2 \theta_{\kk}   \Bigr) \text{Re} \left( q_{\kk} e^{-i(\xi \Lambda_{\kk} + \omega_{0}t)}\right) \\
& \ \  - 2 \gamma_{6}(\sin \theta_{\kk} \cos \nu_{\kk} - 3 \cos \theta_{\kk} \sin \nu_{\kk}) \text{Im} \left( q_{\kk} e^{-i(\xi \Lambda_{\kk} + \omega_{0}t)}\right) \Biggr)
\end{aligned}$ \\
 \hline          
           
  & $1^{\rm st}$     &       $ \begin{aligned}
j^{\rm inter (1)}_{y, \kk} = -\frac{2 e \gamma_{0} a}{\hbar}\Biggl( \xi z_{\kk} \sin(\theta_{\kk} - \nu_{\kk}) \text{Re}\left(q_{\kk} e^{-i(\xi \Lambda_{\kk} + \omega_{0}t)}\right) + \cos \theta_{\kk} \text{Im} \left( q_{\kk} e^{-i(\xi \Lambda_{\kk} + \omega_{0}t)}\right) \Biggr)
\end{aligned}$ \\ 
\multirow{3}{*}{\rot{\rlap{\textbf{$y$ component}}}}  & $2^{\rm nd}$     &      $\begin{aligned}
j^{\rm inter (2)}_{y, \kk} = -\frac{2 e a^2 \ppi}{\hbar}\Biggl( \Biggl( -2 \gamma_{3}  z_{\kk} \sin(2 \theta_{\kk} - \nu_{\kk}) &
 + (\gamma_{1} - \gamma_{2})\frac{g_{\kk}}{\epsilon^{b}_{\kk}} \sin \theta_{\kk} \Biggr) \text{Re} \left( q_{\kk} e^{-i(\xi \Lambda_{\kk} + \omega_{0}t)}\right) \nonumber \\
& - 2 \gamma_{3} \xi \cos(2 \theta_{\kk}-\nu_{\kk}) \text{Im} \left( q_{\kk} e^{-i(\xi \Lambda_{\kk} + \omega_{0}t)}\right) \Biggr)
\end{aligned}
$ \\
 & $3^{\rm rd}$     &     
$\begin{aligned}
j^{\rm inter (3)}_{y, \kk} = -\frac{e a^3 |\ppi|^2}{\hbar} \Bigg( \Bigl( 2 \xi \gamma_{6} z_{\kk}( 2 \sin(\theta_{\kk}- \nu_{\kk}) + \sin(\theta_{\kk} + \nu_{\kk})) - 3 \xi (\gamma_{4} - \gamma_{5}) \frac{g_{\kk}}{\epsilon^{b}_{\kk}} \sin 2 \theta_{\kk} \Bigr)  \text{Re} \left( q_{\kk} e^{-i(\xi \Lambda_{\kk} + \omega_{0}t)}\right)
 \\ 
 + 2 \gamma_{6} (\cos \theta_{\kk} \cos \nu_{\kk} + 3 \sin \theta_{\kk} \sin \nu_{\kk}) \text{Im} \left( q_{\kk} e^{-i(\xi \Lambda_{\kk} + \omega_{0}t)}\right)   
 \Biggr)
\end{aligned}$                          
\end{tabular}ß
\end{ruledtabular}
\caption{\label{tbl:interband_currents_TMD} Contributions to the microscopic current up to third $\kk \cdot p$ order, for both Cartesian components.}
\end{table*}

%%%% bibliography

\clearpage

\twocolumngrid
\nocite{*}
\bibliography{TMD_bib}

%merlin.mbs apsrev4-1.bst 2010-07-25 4.21a (PWD, AO, DPC) hacked
%Control: key (0)
%Control: author (8) initials jnrlst
%Control: editor formatted (1) identically to author
%Control: production of article title (-1) disabled
%Control: page (0) single
%Control: year (1) truncated
%Control: production of eprint (0) enabled
\begin{thebibliography}{31}%
\makeatletter
\providecommand \@ifxundefined [1]{%
 \@ifx{#1\undefined}
}%
\providecommand \@ifnum [1]{%
 \ifnum #1\expandafter \@firstoftwo
 \else \expandafter \@secondoftwo
 \fi
}%
\providecommand \@ifx [1]{%
 \ifx #1\expandafter \@firstoftwo
 \else \expandafter \@secondoftwo
 \fi
}%
\providecommand \natexlab [1]{#1}%
\providecommand \enquote  [1]{``#1''}%
\providecommand \bibnamefont  [1]{#1}%
\providecommand \bibfnamefont [1]{#1}%
\providecommand \citenamefont [1]{#1}%
\providecommand \href@noop [0]{\@secondoftwo}%
\providecommand \href [0]{\begingroup \@sanitize@url \@href}%
\providecommand \@href[1]{\@@startlink{#1}\@@href}%
\providecommand \@@href[1]{\endgroup#1\@@endlink}%
\providecommand \@sanitize@url [0]{\catcode `\\12\catcode `\$12\catcode
  `\&12\catcode `\#12\catcode `\^12\catcode `\_12\catcode `\%12\relax}%
\providecommand \@@startlink[1]{}%
\providecommand \@@endlink[0]{}%
\providecommand \url  [0]{\begingroup\@sanitize@url \@url }%
\providecommand \@url [1]{\endgroup\@href {#1}{\urlprefix }}%
\providecommand \urlprefix  [0]{URL }%
\providecommand \Eprint [0]{\href }%
\providecommand \doibase [0]{http://dx.doi.org/}%
\providecommand \selectlanguage [0]{\@gobble}%
\providecommand \bibinfo  [0]{\@secondoftwo}%
\providecommand \bibfield  [0]{\@secondoftwo}%
\providecommand \translation [1]{[#1]}%
\providecommand \BibitemOpen [0]{}%
\providecommand \bibitemStop [0]{}%
\providecommand \bibitemNoStop [0]{.\EOS\space}%
\providecommand \EOS [0]{\spacefactor3000\relax}%
\providecommand \BibitemShut  [1]{\csname bibitem#1\endcsname}%
\let\auto@bib@innerbib\@empty
%</preamble>
\bibitem [{\citenamefont {Kolobov}\ and\ \citenamefont
  {Tominaga}(2016)}]{Kolobov2016}%
  \BibitemOpen
  \bibfield  {author} {\bibinfo {author} {\bibfnamefont {A.~V.}\ \bibnamefont
  {Kolobov}}\ and\ \bibinfo {author} {\bibfnamefont {J.}~\bibnamefont
  {Tominaga}},\ }\href@noop {} {\emph {\bibinfo {title} {Two-Dimensional
  Transition-Metal Dichalcogenides}}},\ Vol.\ \bibinfo {volume} {239}\
  (\bibinfo  {publisher} {Springer},\ \bibinfo {year} {2016})\BibitemShut
  {NoStop}%
\bibitem [{\citenamefont {Singh}\ \emph {et~al.}(2018)\citenamefont {Singh},
  \citenamefont {Kumar}, \citenamefont {Late}, \citenamefont {Kumar},
  \citenamefont {Patel},\ and\ \citenamefont {Singh}}]{Singh2018}%
  \BibitemOpen
  \bibfield  {author} {\bibinfo {author} {\bibfnamefont {A.~K.}\ \bibnamefont
  {Singh}}, \bibinfo {author} {\bibfnamefont {P.}~\bibnamefont {Kumar}},
  \bibinfo {author} {\bibfnamefont {D.}~\bibnamefont {Late}}, \bibinfo {author}
  {\bibfnamefont {A.}~\bibnamefont {Kumar}}, \bibinfo {author} {\bibfnamefont
  {S.}~\bibnamefont {Patel}}, \ and\ \bibinfo {author} {\bibfnamefont
  {J.}~\bibnamefont {Singh}},\ }\href@noop {} {\bibfield  {journal} {\bibinfo
  {journal} {Applied Materials Today}\ }\textbf {\bibinfo {volume} {13}},\
  \bibinfo {pages} {242} (\bibinfo {year} {2018})}\BibitemShut {NoStop}%
\bibitem [{\citenamefont {Korm\'anyos}\ \emph {et~al.}(2013)\citenamefont
  {Korm\'anyos}, \citenamefont {Z\'olyomi}, \citenamefont {Drummond},
  \citenamefont {Rakyta}, \citenamefont {Burkard},\ and\ \citenamefont
  {Fal'ko}}]{Kormanyos2013}%
  \BibitemOpen
  \bibfield  {author} {\bibinfo {author} {\bibfnamefont {A.}~\bibnamefont
  {Korm\'anyos}}, \bibinfo {author} {\bibfnamefont {V.}~\bibnamefont
  {Z\'olyomi}}, \bibinfo {author} {\bibfnamefont {N.~D.}\ \bibnamefont
  {Drummond}}, \bibinfo {author} {\bibfnamefont {P.}~\bibnamefont {Rakyta}},
  \bibinfo {author} {\bibfnamefont {G.}~\bibnamefont {Burkard}}, \ and\
  \bibinfo {author} {\bibfnamefont {V.~I.}\ \bibnamefont {Fal'ko}},\ }\href
  {\doibase 10.1103/PhysRevB.88.045416} {\bibfield  {journal} {\bibinfo
  {journal} {Phys. Rev. B}\ }\textbf {\bibinfo {volume} {88}},\ \bibinfo
  {pages} {045416} (\bibinfo {year} {2013})}\BibitemShut {NoStop}%
\bibitem [{\citenamefont {S{\"a}yn{\"a}tjoki}\ \emph
  {et~al.}(2017)\citenamefont {S{\"a}yn{\"a}tjoki}, \citenamefont {Karvonen},
  \citenamefont {Rostami}, \citenamefont {Autere}, \citenamefont {Mehravar},
  \citenamefont {Lombardo}, \citenamefont {Norwood}, \citenamefont {Hasan},
  \citenamefont {Peyghambarian}, \citenamefont {Lipsanen} \emph
  {et~al.}}]{Saynatjoki2017}%
  \BibitemOpen
  \bibfield  {author} {\bibinfo {author} {\bibfnamefont {A.}~\bibnamefont
  {S{\"a}yn{\"a}tjoki}}, \bibinfo {author} {\bibfnamefont {L.}~\bibnamefont
  {Karvonen}}, \bibinfo {author} {\bibfnamefont {H.}~\bibnamefont {Rostami}},
  \bibinfo {author} {\bibfnamefont {A.}~\bibnamefont {Autere}}, \bibinfo
  {author} {\bibfnamefont {S.}~\bibnamefont {Mehravar}}, \bibinfo {author}
  {\bibfnamefont {A.}~\bibnamefont {Lombardo}}, \bibinfo {author}
  {\bibfnamefont {R.~A.}\ \bibnamefont {Norwood}}, \bibinfo {author}
  {\bibfnamefont {T.}~\bibnamefont {Hasan}}, \bibinfo {author} {\bibfnamefont
  {N.}~\bibnamefont {Peyghambarian}}, \bibinfo {author} {\bibfnamefont
  {H.}~\bibnamefont {Lipsanen}},  \emph {et~al.},\ }\href@noop {} {\bibfield
  {journal} {\bibinfo  {journal} {Nature communications}\ }\textbf {\bibinfo
  {volume} {8}},\ \bibinfo {pages} {893} (\bibinfo {year} {2017})}\BibitemShut
  {NoStop}%
\bibitem [{\citenamefont {Zhu}\ \emph {et~al.}(2011)\citenamefont {Zhu},
  \citenamefont {Cheng},\ and\ \citenamefont {Schwingenschl\"ogl}}]{Zhu2011}%
  \BibitemOpen
  \bibfield  {author} {\bibinfo {author} {\bibfnamefont {Z.~Y.}\ \bibnamefont
  {Zhu}}, \bibinfo {author} {\bibfnamefont {Y.~C.}\ \bibnamefont {Cheng}}, \
  and\ \bibinfo {author} {\bibfnamefont {U.}~\bibnamefont
  {Schwingenschl\"ogl}},\ }\href {\doibase 10.1103/PhysRevB.84.153402}
  {\bibfield  {journal} {\bibinfo  {journal} {Phys. Rev. B}\ }\textbf {\bibinfo
  {volume} {84}},\ \bibinfo {pages} {153402} (\bibinfo {year}
  {2011})}\BibitemShut {NoStop}%
\bibitem [{\citenamefont {Schaibley}\ \emph {et~al.}(2016)\citenamefont
  {Schaibley}, \citenamefont {Yu}, \citenamefont {Clark}, \citenamefont
  {Rivera}, \citenamefont {Ross}, \citenamefont {Seyler}, \citenamefont {Yao},\
  and\ \citenamefont {Xu}}]{valleytronicsin2d2016}%
  \BibitemOpen
  \bibfield  {author} {\bibinfo {author} {\bibfnamefont {J.~R.}\ \bibnamefont
  {Schaibley}}, \bibinfo {author} {\bibfnamefont {H.}~\bibnamefont {Yu}},
  \bibinfo {author} {\bibfnamefont {G.}~\bibnamefont {Clark}}, \bibinfo
  {author} {\bibfnamefont {P.}~\bibnamefont {Rivera}}, \bibinfo {author}
  {\bibfnamefont {J.~S.}\ \bibnamefont {Ross}}, \bibinfo {author}
  {\bibfnamefont {K.~L.}\ \bibnamefont {Seyler}}, \bibinfo {author}
  {\bibfnamefont {W.}~\bibnamefont {Yao}}, \ and\ \bibinfo {author}
  {\bibfnamefont {X.}~\bibnamefont {Xu}},\ }\href {\doibase 2016/08/23/online}
  {\bibfield  {journal} {\bibinfo  {journal} {Nature Reviews Materials}\
  }\textbf {\bibinfo {volume} {1}} (\bibinfo {year} {2016}),\
  2016/08/23/online}\BibitemShut {NoStop}%
\bibitem [{\citenamefont {Xu}\ \emph {et~al.}(2014)\citenamefont {Xu},
  \citenamefont {Yao}, \citenamefont {Xiao},\ and\ \citenamefont
  {Heinz}}]{Yao2014}%
  \BibitemOpen
  \bibfield  {author} {\bibinfo {author} {\bibfnamefont {X.}~\bibnamefont
  {Xu}}, \bibinfo {author} {\bibfnamefont {W.}~\bibnamefont {Yao}}, \bibinfo
  {author} {\bibfnamefont {D.}~\bibnamefont {Xiao}}, \ and\ \bibinfo {author}
  {\bibfnamefont {T.~F.}\ \bibnamefont {Heinz}},\ }\href {\doibase
  10.1038/nphys2942} {\bibfield  {journal} {\bibinfo  {journal} {Nature
  Physics}\ }\textbf {\bibinfo {volume} {10}},\ \bibinfo {pages} {343–350}
  (\bibinfo {year} {2014})}\BibitemShut {NoStop}%
\bibitem [{\citenamefont {Xiao}\ \emph {et~al.}(2012)\citenamefont {Xiao},
  \citenamefont {Liu}, \citenamefont {Feng}, \citenamefont {Xu},\ and\
  \citenamefont {Yao}}]{Xiao2012}%
  \BibitemOpen
  \bibfield  {author} {\bibinfo {author} {\bibfnamefont {D.}~\bibnamefont
  {Xiao}}, \bibinfo {author} {\bibfnamefont {G.-B.}\ \bibnamefont {Liu}},
  \bibinfo {author} {\bibfnamefont {W.}~\bibnamefont {Feng}}, \bibinfo {author}
  {\bibfnamefont {X.}~\bibnamefont {Xu}}, \ and\ \bibinfo {author}
  {\bibfnamefont {W.}~\bibnamefont {Yao}},\ }\href@noop {} {\bibfield
  {journal} {\bibinfo  {journal} {Physical Review Letters}\ }\textbf {\bibinfo
  {volume} {108}},\ \bibinfo {pages} {196802} (\bibinfo {year}
  {2012})}\BibitemShut {NoStop}%
\bibitem [{\citenamefont {Zheng}\ \emph {et~al.}(2018)\citenamefont {Zheng},
  \citenamefont {Jiang}, \citenamefont {Hu}, \citenamefont {Li}, \citenamefont
  {Zeng}, \citenamefont {Wang},\ and\ \citenamefont {Pan}}]{Zheng2018}%
  \BibitemOpen
  \bibfield  {author} {\bibinfo {author} {\bibfnamefont {W.}~\bibnamefont
  {Zheng}}, \bibinfo {author} {\bibfnamefont {Y.}~\bibnamefont {Jiang}},
  \bibinfo {author} {\bibfnamefont {X.}~\bibnamefont {Hu}}, \bibinfo {author}
  {\bibfnamefont {H.}~\bibnamefont {Li}}, \bibinfo {author} {\bibfnamefont
  {Z.}~\bibnamefont {Zeng}}, \bibinfo {author} {\bibfnamefont {X.}~\bibnamefont
  {Wang}}, \ and\ \bibinfo {author} {\bibfnamefont {A.}~\bibnamefont {Pan}},\
  }\href@noop {} {\bibfield  {journal} {\bibinfo  {journal} {Advanced Optical
  Materials}\ ,\ \bibinfo {pages} {1800420}} (\bibinfo {year}
  {2018})}\BibitemShut {NoStop}%
\bibitem [{\citenamefont {Liu}\ \emph {et~al.}(2016)\citenamefont {Liu},
  \citenamefont {Li}, \citenamefont {You}, \citenamefont {Ghimire},
  \citenamefont {Heinz},\ and\ \citenamefont {Reis}}]{Liu2016}%
  \BibitemOpen
  \bibfield  {author} {\bibinfo {author} {\bibfnamefont {H.}~\bibnamefont
  {Liu}}, \bibinfo {author} {\bibfnamefont {Y.}~\bibnamefont {Li}}, \bibinfo
  {author} {\bibfnamefont {Y.~S.}\ \bibnamefont {You}}, \bibinfo {author}
  {\bibfnamefont {S.}~\bibnamefont {Ghimire}}, \bibinfo {author} {\bibfnamefont
  {T.~F.}\ \bibnamefont {Heinz}}, \ and\ \bibinfo {author} {\bibfnamefont
  {D.~A.}\ \bibnamefont {Reis}},\ }\href@noop {} {\bibfield  {journal}
  {\bibinfo  {journal} {Nature Physics}\ } (\bibinfo {year}
  {2016})}\BibitemShut {NoStop}%
\bibitem [{\citenamefont {Yu}\ \emph {et~al.}(2014)\citenamefont {Yu},
  \citenamefont {Liu}, \citenamefont {Gong}, \citenamefont {Xu},\ and\
  \citenamefont {Yao}}]{Yu2014}%
  \BibitemOpen
  \bibfield  {author} {\bibinfo {author} {\bibfnamefont {H.}~\bibnamefont
  {Yu}}, \bibinfo {author} {\bibfnamefont {G.-B.}\ \bibnamefont {Liu}},
  \bibinfo {author} {\bibfnamefont {P.}~\bibnamefont {Gong}}, \bibinfo {author}
  {\bibfnamefont {X.}~\bibnamefont {Xu}}, \ and\ \bibinfo {author}
  {\bibfnamefont {W.}~\bibnamefont {Yao}},\ }\href@noop {} {\bibfield
  {journal} {\bibinfo  {journal} {Nature communications}\ }\textbf {\bibinfo
  {volume} {5}} (\bibinfo {year} {2014})}\BibitemShut {NoStop}%
\bibitem [{\citenamefont {Autere}\ \emph {et~al.}(2018)\citenamefont {Autere},
  \citenamefont {Jussila}, \citenamefont {Marini}, \citenamefont {Saavedra},
  \citenamefont {Dai}, \citenamefont {S\"ayn\"atjoki}, \citenamefont
  {Karvonen}, \citenamefont {Yang}, \citenamefont {Amirsolaimani},
  \citenamefont {Norwood}, \citenamefont {Peyghambarian}, \citenamefont
  {Lipsanen}, \citenamefont {Kieu}, \citenamefont {de~Abajo},\ and\
  \citenamefont {Sun}}]{Marini2018}%
  \BibitemOpen
  \bibfield  {author} {\bibinfo {author} {\bibfnamefont {A.}~\bibnamefont
  {Autere}}, \bibinfo {author} {\bibfnamefont {H.}~\bibnamefont {Jussila}},
  \bibinfo {author} {\bibfnamefont {A.}~\bibnamefont {Marini}}, \bibinfo
  {author} {\bibfnamefont {J.~R.~M.}\ \bibnamefont {Saavedra}}, \bibinfo
  {author} {\bibfnamefont {Y.}~\bibnamefont {Dai}}, \bibinfo {author}
  {\bibfnamefont {A.}~\bibnamefont {S\"ayn\"atjoki}}, \bibinfo {author}
  {\bibfnamefont {L.}~\bibnamefont {Karvonen}}, \bibinfo {author}
  {\bibfnamefont {H.}~\bibnamefont {Yang}}, \bibinfo {author} {\bibfnamefont
  {B.}~\bibnamefont {Amirsolaimani}}, \bibinfo {author} {\bibfnamefont {R.~A.}\
  \bibnamefont {Norwood}}, \bibinfo {author} {\bibfnamefont {N.}~\bibnamefont
  {Peyghambarian}}, \bibinfo {author} {\bibfnamefont {H.}~\bibnamefont
  {Lipsanen}}, \bibinfo {author} {\bibfnamefont {K.}~\bibnamefont {Kieu}},
  \bibinfo {author} {\bibfnamefont {F.~J.~G.}\ \bibnamefont {de~Abajo}}, \ and\
  \bibinfo {author} {\bibfnamefont {Z.}~\bibnamefont {Sun}},\ }\href {\doibase
  10.1103/PhysRevB.98.115426} {\bibfield  {journal} {\bibinfo  {journal} {Phys.
  Rev. B}\ }\textbf {\bibinfo {volume} {98}},\ \bibinfo {pages} {115426}
  (\bibinfo {year} {2018})}\BibitemShut {NoStop}%
\bibitem [{\citenamefont {Khorasani}(2018)}]{Khorasani2018}%
  \BibitemOpen
  \bibfield  {author} {\bibinfo {author} {\bibfnamefont {S.}~\bibnamefont
  {Khorasani}},\ }\href@noop {} {\bibfield  {journal} {\bibinfo  {journal}
  {Communications in Theoretical Physics}\ }\textbf {\bibinfo {volume} {70}},\
  \bibinfo {pages} {344} (\bibinfo {year} {2018})}\BibitemShut {NoStop}%
\bibitem [{\citenamefont {Wang}\ \emph {et~al.}(2018)\citenamefont {Wang},
  \citenamefont {Guo}, \citenamefont {You}, \citenamefont {Zhang},
  \citenamefont {Zheng},\ and\ \citenamefont {Cheng}}]{Wang2018}%
  \BibitemOpen
  \bibfield  {author} {\bibinfo {author} {\bibfnamefont {Y.}~\bibnamefont
  {Wang}}, \bibinfo {author} {\bibfnamefont {Z.}~\bibnamefont {Guo}}, \bibinfo
  {author} {\bibfnamefont {J.}~\bibnamefont {You}}, \bibinfo {author}
  {\bibfnamefont {Z.}~\bibnamefont {Zhang}}, \bibinfo {author} {\bibfnamefont
  {X.}~\bibnamefont {Zheng}}, \ and\ \bibinfo {author} {\bibfnamefont
  {X.}~\bibnamefont {Cheng}},\ }\href@noop {} {\bibfield  {journal} {\bibinfo
  {journal} {Photonic Sensors}\ ,\ \bibinfo {pages} {1}} (\bibinfo {year}
  {2018})}\BibitemShut {NoStop}%
\bibitem [{\citenamefont {Xia}\ \emph {et~al.}(2014)\citenamefont {Xia},
  \citenamefont {Wang}, \citenamefont {Xiao}, \citenamefont {Dubey},\ and\
  \citenamefont {Ramasubramaniam}}]{Xia2014}%
  \BibitemOpen
  \bibfield  {author} {\bibinfo {author} {\bibfnamefont {F.}~\bibnamefont
  {Xia}}, \bibinfo {author} {\bibfnamefont {H.}~\bibnamefont {Wang}}, \bibinfo
  {author} {\bibfnamefont {D.}~\bibnamefont {Xiao}}, \bibinfo {author}
  {\bibfnamefont {M.}~\bibnamefont {Dubey}}, \ and\ \bibinfo {author}
  {\bibfnamefont {A.}~\bibnamefont {Ramasubramaniam}},\ }\href@noop {}
  {\bibfield  {journal} {\bibinfo  {journal} {Nature Photonics}\ }\textbf
  {\bibinfo {volume} {8}},\ \bibinfo {pages} {899} (\bibinfo {year}
  {2014})}\BibitemShut {NoStop}%
\bibitem [{\citenamefont {Gmitra}\ \emph {et~al.}(2016)\citenamefont {Gmitra},
  \citenamefont {Kochan}, \citenamefont {H{\"o}gl},\ and\ \citenamefont
  {Fabian}}]{Gmitra2016}%
  \BibitemOpen
  \bibfield  {author} {\bibinfo {author} {\bibfnamefont {M.}~\bibnamefont
  {Gmitra}}, \bibinfo {author} {\bibfnamefont {D.}~\bibnamefont {Kochan}},
  \bibinfo {author} {\bibfnamefont {P.}~\bibnamefont {H{\"o}gl}}, \ and\
  \bibinfo {author} {\bibfnamefont {J.}~\bibnamefont {Fabian}},\ }\href@noop {}
  {\bibfield  {journal} {\bibinfo  {journal} {Physical Review B}\ }\textbf
  {\bibinfo {volume} {93}},\ \bibinfo {pages} {155104} (\bibinfo {year}
  {2016})}\BibitemShut {NoStop}%
\bibitem [{\citenamefont {Liu}\ \emph {et~al.}(2013)\citenamefont {Liu},
  \citenamefont {Shan}, \citenamefont {Yao}, \citenamefont {Yao},\ and\
  \citenamefont {Xiao}}]{Liu2013}%
  \BibitemOpen
  \bibfield  {author} {\bibinfo {author} {\bibfnamefont {G.-B.}\ \bibnamefont
  {Liu}}, \bibinfo {author} {\bibfnamefont {W.-Y.}\ \bibnamefont {Shan}},
  \bibinfo {author} {\bibfnamefont {Y.}~\bibnamefont {Yao}}, \bibinfo {author}
  {\bibfnamefont {W.}~\bibnamefont {Yao}}, \ and\ \bibinfo {author}
  {\bibfnamefont {D.}~\bibnamefont {Xiao}},\ }\href@noop {} {\bibfield
  {journal} {\bibinfo  {journal} {Physical Review B}\ }\textbf {\bibinfo
  {volume} {88}},\ \bibinfo {pages} {085433} (\bibinfo {year}
  {2013})}\BibitemShut {NoStop}%
\bibitem [{\citenamefont {Rostami}\ \emph {et~al.}(2013)\citenamefont
  {Rostami}, \citenamefont {Moghaddam},\ and\ \citenamefont
  {Asgari}}]{Rostami2013}%
  \BibitemOpen
  \bibfield  {author} {\bibinfo {author} {\bibfnamefont {H.}~\bibnamefont
  {Rostami}}, \bibinfo {author} {\bibfnamefont {A.~G.}\ \bibnamefont
  {Moghaddam}}, \ and\ \bibinfo {author} {\bibfnamefont {R.}~\bibnamefont
  {Asgari}},\ }\href@noop {} {\bibfield  {journal} {\bibinfo  {journal}
  {Physical Review B}\ }\textbf {\bibinfo {volume} {88}},\ \bibinfo {pages}
  {085440} (\bibinfo {year} {2013})}\BibitemShut {NoStop}%
\bibitem [{\citenamefont {Carvalho}\ \emph {et~al.}(2018)\citenamefont
  {Carvalho}, \citenamefont {Biancalana},\ and\ \citenamefont
  {Marini}}]{Carvalho2018}%
  \BibitemOpen
  \bibfield  {author} {\bibinfo {author} {\bibfnamefont {D.~N.}\ \bibnamefont
  {Carvalho}}, \bibinfo {author} {\bibfnamefont {F.}~\bibnamefont
  {Biancalana}}, \ and\ \bibinfo {author} {\bibfnamefont {A.}~\bibnamefont
  {Marini}},\ }\href@noop {} {\bibfield  {journal} {\bibinfo  {journal}
  {Physical Review B}\ }\textbf {\bibinfo {volume} {97}},\ \bibinfo {pages}
  {195123} (\bibinfo {year} {2018})}\BibitemShut {NoStop}%
\bibitem [{\citenamefont {Voon}\ and\ \citenamefont
  {Willatzen}(2009)}]{Voon2009}%
  \BibitemOpen
  \bibfield  {author} {\bibinfo {author} {\bibfnamefont {L.~C. L.~Y.}\
  \bibnamefont {Voon}}\ and\ \bibinfo {author} {\bibfnamefont {M.}~\bibnamefont
  {Willatzen}},\ }\href@noop {} {\emph {\bibinfo {title} {The kp method:
  electronic properties of semiconductors}}}\ (\bibinfo  {publisher} {Springer
  Science \& Business Media},\ \bibinfo {year} {2009})\BibitemShut {NoStop}%
\bibitem [{\citenamefont {Zahid}\ \emph {et~al.}(2013)\citenamefont {Zahid},
  \citenamefont {Liu}, \citenamefont {Zhu}, \citenamefont {Wang},\ and\
  \citenamefont {Guo}}]{Zahid2013}%
  \BibitemOpen
  \bibfield  {author} {\bibinfo {author} {\bibfnamefont {F.}~\bibnamefont
  {Zahid}}, \bibinfo {author} {\bibfnamefont {L.}~\bibnamefont {Liu}}, \bibinfo
  {author} {\bibfnamefont {Y.}~\bibnamefont {Zhu}}, \bibinfo {author}
  {\bibfnamefont {J.}~\bibnamefont {Wang}}, \ and\ \bibinfo {author}
  {\bibfnamefont {H.}~\bibnamefont {Guo}},\ }\href@noop {} {\bibfield
  {journal} {\bibinfo  {journal} {Aip Advances}\ }\textbf {\bibinfo {volume}
  {3}},\ \bibinfo {pages} {052111} (\bibinfo {year} {2013})}\BibitemShut
  {NoStop}%
\bibitem [{\citenamefont {Alidoust}\ \emph {et~al.}(2014)\citenamefont
  {Alidoust}, \citenamefont {Bian}, \citenamefont {Xu}, \citenamefont {Sankar},
  \citenamefont {Neupane}, \citenamefont {Liu}, \citenamefont {Belopolski},
  \citenamefont {Qu}, \citenamefont {Denlinger}, \citenamefont {Chou} \emph
  {et~al.}}]{Alidoust2014}%
  \BibitemOpen
  \bibfield  {author} {\bibinfo {author} {\bibfnamefont {N.}~\bibnamefont
  {Alidoust}}, \bibinfo {author} {\bibfnamefont {G.}~\bibnamefont {Bian}},
  \bibinfo {author} {\bibfnamefont {S.-Y.}\ \bibnamefont {Xu}}, \bibinfo
  {author} {\bibfnamefont {R.}~\bibnamefont {Sankar}}, \bibinfo {author}
  {\bibfnamefont {M.}~\bibnamefont {Neupane}}, \bibinfo {author} {\bibfnamefont
  {C.}~\bibnamefont {Liu}}, \bibinfo {author} {\bibfnamefont {I.}~\bibnamefont
  {Belopolski}}, \bibinfo {author} {\bibfnamefont {D.-X.}\ \bibnamefont {Qu}},
  \bibinfo {author} {\bibfnamefont {J.~D.}\ \bibnamefont {Denlinger}}, \bibinfo
  {author} {\bibfnamefont {F.-C.}\ \bibnamefont {Chou}},  \emph {et~al.},\
  }\href@noop {} {\bibfield  {journal} {\bibinfo  {journal} {Nature
  communications}\ }\textbf {\bibinfo {volume} {5}},\ \bibinfo {pages} {1}
  (\bibinfo {year} {2014})}\BibitemShut {NoStop}%
\bibitem [{\citenamefont {Ishikawa}(2010)}]{Ishikawa2010}%
  \BibitemOpen
  \bibfield  {author} {\bibinfo {author} {\bibfnamefont {K.~L.}\ \bibnamefont
  {Ishikawa}},\ }\href@noop {} {\bibfield  {journal} {\bibinfo  {journal}
  {Physical Review B}\ }\textbf {\bibinfo {volume} {82}},\ \bibinfo {pages}
  {201402} (\bibinfo {year} {2010})}\BibitemShut {NoStop}%
\bibitem [{\citenamefont {Carvalho}\ \emph {et~al.}(2017)\citenamefont
  {Carvalho}, \citenamefont {Marini},\ and\ \citenamefont
  {Biancalana}}]{Carvalho2017}%
  \BibitemOpen
  \bibfield  {author} {\bibinfo {author} {\bibfnamefont {D.~N.}\ \bibnamefont
  {Carvalho}}, \bibinfo {author} {\bibfnamefont {A.}~\bibnamefont {Marini}}, \
  and\ \bibinfo {author} {\bibfnamefont {F.}~\bibnamefont {Biancalana}},\
  }\href@noop {} {\bibfield  {journal} {\bibinfo  {journal} {Annals of
  Physics}\ }\textbf {\bibinfo {volume} {378}},\ \bibinfo {pages} {24}
  (\bibinfo {year} {2017})}\BibitemShut {NoStop}%
\bibitem [{\citenamefont {McCann}\ and\ \citenamefont
  {Koshino}(2013)}]{Mccann2013}%
  \BibitemOpen
  \bibfield  {author} {\bibinfo {author} {\bibfnamefont {E.}~\bibnamefont
  {McCann}}\ and\ \bibinfo {author} {\bibfnamefont {M.}~\bibnamefont
  {Koshino}},\ }\href@noop {} {\bibfield  {journal} {\bibinfo  {journal}
  {Reports on Progress in Physics}\ }\textbf {\bibinfo {volume} {76}},\
  \bibinfo {pages} {056503} (\bibinfo {year} {2013})}\BibitemShut {NoStop}%
\bibitem [{\citenamefont {Liu}\ \emph {et~al.}(2010)\citenamefont {Liu},
  \citenamefont {Qi}, \citenamefont {Zhang}, \citenamefont {Dai}, \citenamefont
  {Fang},\ and\ \citenamefont {Zhang}}]{Liu2010}%
  \BibitemOpen
  \bibfield  {author} {\bibinfo {author} {\bibfnamefont {C.-X.}\ \bibnamefont
  {Liu}}, \bibinfo {author} {\bibfnamefont {X.-L.}\ \bibnamefont {Qi}},
  \bibinfo {author} {\bibfnamefont {H.}~\bibnamefont {Zhang}}, \bibinfo
  {author} {\bibfnamefont {X.}~\bibnamefont {Dai}}, \bibinfo {author}
  {\bibfnamefont {Z.}~\bibnamefont {Fang}}, \ and\ \bibinfo {author}
  {\bibfnamefont {S.-C.}\ \bibnamefont {Zhang}},\ }\href@noop {} {\bibfield
  {journal} {\bibinfo  {journal} {Physical Review B}\ }\textbf {\bibinfo
  {volume} {82}},\ \bibinfo {pages} {045122} (\bibinfo {year}
  {2010})}\BibitemShut {NoStop}%
\bibitem [{\citenamefont {Villari}\ \emph {et~al.}(2018)\citenamefont
  {Villari}, \citenamefont {Galbraith},\ and\ \citenamefont
  {Biancalana}}]{DiMauro2018}%
  \BibitemOpen
  \bibfield  {author} {\bibinfo {author} {\bibfnamefont {L.~D.~M.}\
  \bibnamefont {Villari}}, \bibinfo {author} {\bibfnamefont {I.}~\bibnamefont
  {Galbraith}}, \ and\ \bibinfo {author} {\bibfnamefont {F.}~\bibnamefont
  {Biancalana}},\ }\href {\doibase 10.1103/PhysRevB.98.205402} {\bibfield
  {journal} {\bibinfo  {journal} {Phys. Rev. B}\ }\textbf {\bibinfo {volume}
  {98}},\ \bibinfo {pages} {205402} (\bibinfo {year} {2018})}\BibitemShut
  {NoStop}%
\bibitem [{\citenamefont {Van't~Erve}\ \emph {et~al.}(2016)\citenamefont
  {Van't~Erve}, \citenamefont {Hanbicki}, \citenamefont {Friedman},
  \citenamefont {McCreary}, \citenamefont {Cobas}, \citenamefont {Li},
  \citenamefont {Robinson},\ and\ \citenamefont {Jonker}}]{vanErve2016}%
  \BibitemOpen
  \bibfield  {author} {\bibinfo {author} {\bibfnamefont {O.~M.}\ \bibnamefont
  {Van't~Erve}}, \bibinfo {author} {\bibfnamefont {A.~T.}\ \bibnamefont
  {Hanbicki}}, \bibinfo {author} {\bibfnamefont {A.~L.}\ \bibnamefont
  {Friedman}}, \bibinfo {author} {\bibfnamefont {K.~M.}\ \bibnamefont
  {McCreary}}, \bibinfo {author} {\bibfnamefont {E.}~\bibnamefont {Cobas}},
  \bibinfo {author} {\bibfnamefont {C.~H.}\ \bibnamefont {Li}}, \bibinfo
  {author} {\bibfnamefont {J.~T.}\ \bibnamefont {Robinson}}, \ and\ \bibinfo
  {author} {\bibfnamefont {B.~T.}\ \bibnamefont {Jonker}},\ }\href@noop {}
  {\bibfield  {journal} {\bibinfo  {journal} {Journal of Materials Research}\
  }\textbf {\bibinfo {volume} {31}},\ \bibinfo {pages} {845} (\bibinfo {year}
  {2016})}\BibitemShut {NoStop}%
\bibitem [{\citenamefont {Dey}\ \emph {et~al.}(2016)\citenamefont {Dey},
  \citenamefont {Paul}, \citenamefont {Wang}, \citenamefont {Stevens},
  \citenamefont {Liu}, \citenamefont {Romero}, \citenamefont {Shan},
  \citenamefont {Hilton},\ and\ \citenamefont {Karaiskaj}}]{Dey2016}%
  \BibitemOpen
  \bibfield  {author} {\bibinfo {author} {\bibfnamefont {P.}~\bibnamefont
  {Dey}}, \bibinfo {author} {\bibfnamefont {J.}~\bibnamefont {Paul}}, \bibinfo
  {author} {\bibfnamefont {Z.}~\bibnamefont {Wang}}, \bibinfo {author}
  {\bibfnamefont {C.}~\bibnamefont {Stevens}}, \bibinfo {author} {\bibfnamefont
  {C.}~\bibnamefont {Liu}}, \bibinfo {author} {\bibfnamefont {A.}~\bibnamefont
  {Romero}}, \bibinfo {author} {\bibfnamefont {J.}~\bibnamefont {Shan}},
  \bibinfo {author} {\bibfnamefont {D.}~\bibnamefont {Hilton}}, \ and\ \bibinfo
  {author} {\bibfnamefont {D.}~\bibnamefont {Karaiskaj}},\ }\href@noop {}
  {\bibfield  {journal} {\bibinfo  {journal} {Physical review letters}\
  }\textbf {\bibinfo {volume} {116}},\ \bibinfo {pages} {127402} (\bibinfo
  {year} {2016})}\BibitemShut {NoStop}%
\bibitem [{\citenamefont {Mak}\ and\ \citenamefont {Shan}(2016)}]{Mak2016}%
  \BibitemOpen
  \bibfield  {author} {\bibinfo {author} {\bibfnamefont {K.~F.}\ \bibnamefont
  {Mak}}\ and\ \bibinfo {author} {\bibfnamefont {J.}~\bibnamefont {Shan}},\
  }\href@noop {} {\bibfield  {journal} {\bibinfo  {journal} {Nature Photonics}\
  }\textbf {\bibinfo {volume} {10}},\ \bibinfo {pages} {216} (\bibinfo {year}
  {2016})}\BibitemShut {NoStop}%
\bibitem [{\citenamefont {Wang}(2013)}]{WangBook}%
  \BibitemOpen
  \bibfield  {author} {\bibinfo {author} {\bibfnamefont {Z.~M.}\ \bibnamefont
  {Wang}},\ }\href@noop {} {\emph {\bibinfo {title} {MoS2: materials, physics,
  and devices}}},\ Vol.~\bibinfo {volume} {21}\ (\bibinfo  {publisher}
  {Springer Science \& Business Media},\ \bibinfo {year} {2013})\BibitemShut
  {NoStop}%
\end{thebibliography}%

\end{document}